\documentstyle[12pt,indent,epsf,eqsection,subeqnarray]{article}

\begin{document}

\bibliographystyle{plain}

\title{Logarithmic Corrections and Finite-Size Scaling in the 
       Two-Dimensional 4-State Potts Model}  
\author{
  {\small Jes\'us Salas}                  \\[-0.2cm]
  {\small Alan D.~Sokal}                  \\[-0.2cm]
  {\small\it Department of Physics}       \\[-0.2cm]
  {\small\it New York University}         \\[-0.2cm]
  {\small\it 4 Washington Place}          \\[-0.2cm]
  {\small\it New York, NY 10003 USA}      \\[-0.2cm]
  {\small\tt SALAS@MAFALDA.PHYSICS.NYU.EDU},
                    {\small\tt SOKAL@NYU.EDU}   \\[-0.2cm]
  {\protect\makebox[5in]{\quad}}  
  \\
}
\vspace{0.5cm}

\maketitle
\thispagestyle{empty}   

\def\spose#1{\hbox to 0pt{#1\hss}}
\def\ltapprox{\mathrel{\spose{\lower 3pt\hbox{$\mathchar"218$}}
 \raise 2.0pt\hbox{$\mathchar"13C$}}}
\def\gtapprox{\mathrel{\spose{\lower 3pt\hbox{$\mathchar"218$}}
 \raise 2.0pt\hbox{$\mathchar"13E$}}}
\def\inapprox{\mathrel{\spose{\lower 3pt\hbox{$\mathchar"218$}}
 \raise 2.0pt\hbox{$\mathchar"232$}}}

\begin{abstract}
We analyze the scaling and finite-size-scaling 
behavior of the two-dimensional 4-state 
Potts model. We find new multiplicative logarithmic corrections for the 
susceptibility, in addition to the already known ones for the specific heat.   
We also find additive logarithmic corrections to scaling,
some of which are universal.
We have checked the theoretical predictions at criticality and off criticality 
by means of high-precision Monte Carlo data.   
\end{abstract}

\bigskip 
\noindent 
{\bf Key Words:} Scaling; finite-size scaling; critical phenomena;
renormalization group; logarithmic corrections; corrections to scaling;
Potts model; Ashkin-Teller model; Swendsen-Wang algorithm;  
cluster algorithm; Monte Carlo.

\clearpage

\newcommand{\be}{\begin{equation}}
\newcommand{\ee}{\end{equation}}
\newcommand{\<}{\langle}
\renewcommand{\>}{\rangle}
\newcommand{\para}{\|}
\renewcommand{\perp}{\bot}

\def\smfrac#1#2{{\textstyle\frac{#1}{#2}}}
\def\half{ {{1 \over 2 }}}
\def\smhalf{ {\smfrac{1}{2}} }
\def\scra{{\cal A}}
\def\scrc{{\cal C}}
\def\scrd{{\cal D}}
\def\scre{{\cal E}}
\def\scrf{{\cal F}}
\def\scrh{{\cal H}}
\def\scrk{{\cal K}}
\def\scrm{{\cal M}}
\newcommand{\scrmvec}{\vec{\cal M}_V}
\def\scrmtens{{\stackrel{\leftrightarrow}{\cal M}_T}}
\def\scro{{\cal O}}
\def\scrp{{\cal P}}
\def\scrr{{\cal R}}
\def\scrs{{\cal S}}
\def\ttens{{\stackrel{\leftrightarrow}{T}}}
\def\scrv{{\cal V}}
\def\scrw{{\cal W}}
\def\scry{{\cal Y}}
\def\tauss{\tau_{int,\,\scrm^2}}
\def\taux{\tau_{int,\,{\cal M}^2}}
\newcommand{\taum}{\tau_{int,\,\vec{\cal M}}}
\def\taue{\tau_{int,\,{\cal E}}}
\newcommand{\imag}{\mathop{\rm Im}\nolimits}
\newcommand{\real}{\mathop{\rm Re}\nolimits}
\newcommand{\tr}{\mathop{\rm tr}\nolimits}
\newcommand{\sgn}{\mathop{\rm sgn}\nolimits}
\newcommand{\codim}{\mathop{\rm codim}\nolimits}
\newcommand{\rank}{\mathop{\rm rank}\nolimits}
\newcommand{\sech}{\mathop{\rm sech}\nolimits}
\def\textprime{{${}^\prime$}}
\newcommand{\longto}{\longrightarrow}
\def\var{ \hbox{var} }
\newcommand{\gtilde}{ {\widetilde{G}} }
\newcommand{\USp}{ \hbox{\it USp} }
\newcommand{\CP}{ \hbox{\it CP\/} }
\newcommand{\QP}{ \hbox{\it QP\/} }
\def\hboxscript#1{ {\hbox{\scriptsize\em #1}} }

\newcommand{\plotdot}{\makebox(0,0){$\bullet$}}
\newcommand{\plotsmalldot}{\makebox(0,0){{\footnotesize $\bullet$}}}

\def\bsigma{\mbox{\protect\boldmath $\sigma$}}
\def\bpi{\mbox{\protect\boldmath $\pi$}}
\def\btau{\mbox{\protect\boldmath $\tau$}}
\def\bn{{\bf n}}
\def\br{{\bf r}}
\def\bz{{\bf z}}
\def\bh{\mbox{\protect\boldmath $h$}}

\def\betatilde{ {\widetilde{\beta}} }
\def\hatp{\hat p}
\def\hatl{\hat l}

\def\msbar{ {\overline{\hbox{\scriptsize MS}}} }
\def\normalmsbar{ {\overline{\hbox{\normalsize MS}}} }

\def\eff{ {\hbox{\scriptsize\em eff}} }

\newcommand{\reff}[1]{(\ref{#1})}

\newcommand{\Z}{{\bf Z}}
\newcommand{\zed}{{\bf \Z}}
\newcommand{\R}{\hbox{{\rm I}\kern-.2em\hbox{\rm R}}}
\font\srm=cmr7 		
\def\szed{\hbox{\srm Z\kern-.45em\hbox{\srm Z}}}
\def\sR{\hbox{{\srm I}\kern-.2em\hbox{\srm R}}}
\def\C{{\bf C}}



\newtheorem{theorem}{Theorem}[section]
\newtheorem{corollary}[theorem]{Corollary}
\newtheorem{lemma}[theorem]{Lemma}
\newtheorem{conjecture}[theorem]{Conjecture}
\newtheorem{definition}[theorem]{Definition}
\def\proof{\bigskip\par\noindent{\sc Proof.\ }}
\def\qed{\hbox{\hskip 6pt\vrule width6pt height7pt depth1pt \hskip1pt}\bigskip}

%
%
\newenvironment{sarray}{
          \textfont0=\scriptfont0
          \scriptfont0=\scriptscriptfont0
          \textfont1=\scriptfont1
          \scriptfont1=\scriptscriptfont1
          \textfont2=\scriptfont2
          \scriptfont2=\scriptscriptfont2
          \textfont3=\scriptfont3
          \scriptfont3=\scriptscriptfont3
        \renewcommand{\arraystretch}{0.7}
        \begin{array}{l}}{\end{array}}

\newenvironment{scarray}{
          \textfont0=\scriptfont0
          \scriptfont0=\scriptscriptfont0
          \textfont1=\scriptfont1
          \scriptfont1=\scriptscriptfont1
          \textfont2=\scriptfont2
          \scriptfont2=\scriptscriptfont2
          \textfont3=\scriptfont3
          \scriptfont3=\scriptscriptfont3
        \renewcommand{\arraystretch}{0.7}
        \begin{array}{c}}{\end{array}}

%
%
\section{Introduction}  \label{sec_intro}

The two-dimensional (2D) 
$q$-state Potts model is one of the most intensively studied systems in 
statistical mechanics. Although this model is very simple to formulate, 
it has a rich phase diagram.  
Baxter \cite{Baxter_73} showed that for $q\le 4$ this model undergoes a 
second-order phase transition at the self-dual point,
while for $q > 4$ the transition is of first order.

The borderline case $q=4$ is the most difficult:
here the transition is second-order,
but the leading power-law scaling behavior is modified by
multiplicative logarithms,
as was first observed by Nauenberg and Scalapino \cite{Nauenberg_80} and 
Cardy, Nauenberg and Scalapino \cite{Cardy_80}.\footnote{
   In addition to these analytical works,
   the possible existence of multiplicative logarithmic corrections
   was proposed almost contemporaneously by
   Rebbi and Swendsen \cite{Rebbi_80}
   on the basis of numerical Monte Carlo renormalization group data.
}
These authors studied an extended Potts model
(the dilute Potts model \cite{Nienhuis_79})
in which, in addition to the usual $q$ spin states, vacancies are allowed.
The density of the vacancies is governed by a fugacity (``dilution field'').
For each $q<4$, one finds the following phase diagram: 
for low vacancy fugacity, there is an ordinary critical point belonging to the 
universality class of the pure Potts model; for high vacancy fugacity,
there is a first-order transition at which the energy, the magnetization and
the vacancy density are all discontinuous; these two transition curves meet
at a tricritical point. 
In the renormalization-group (RG) framework, there is for each $q<4$ an 
ordinary critical fixed point, a tricritical fixed point, and a 
discontinuity fixed point (governing the first-order transition).
At $q=4$, the critical and tricritical fixed points merge, and the dilution 
field becomes marginal: that is the cause of the multiplicative 
logarithmic corrections. For $q>4$, only the discontinuity fixed point 
survives.

The quantitative analysis of Cardy {\em et al.}\/ \cite{Nauenberg_80,Cardy_80}
is based on studying the RG flow for the $q$-state dilute Potts model,
as a function of the continuous parameter $q$, in a neighborhood of $q=4$.
First, by using a nonlinear transformation to appropriate scaling fields,
they obtain the RG equations in Poincar\'e normal form
through second order in the scaling fields
and first order in $\epsilon \equiv q-4$.
Next they fix the free parameters in this normal form
by matching the RG predictions to leading order in $\epsilon$
with the exactly known results for
the critical exponents when $q<4$ \cite{denNijs_79,Nienhuis_80,Pearson_80}
and for the latent heat when $q>4$ \cite{Baxter_73}.\footnote{
   At that time that Refs.~\cite{Nauenberg_80,Cardy_80}
   were written, the exact formulae for the critical exponents
   were only conjectured.
   They have now been confirmed by both Coulomb-gas and conformal-field-theory
   methods:  see Appendix \ref{secA.1} for a brief review.
}
Finally, by integrating the RG equations at $q=4$
with suitable boundary conditions, they obtain predictions for the 
leading-order critical behavior of:
a) the specific heat, the correlation length, 
and the magnetization as functions of the 
thermal field at zero ordering field;  b) the critical magnetization as a 
function of the ordering field;  and c) the critical two-point correlator as 
a function of the distance between the two points.  
In all these cases, the power-law dependence expected generically
for a second-order phase 
transition is modified by multiplicative logarithmic corrections. 

In this paper we show that by extending the analysis of
Cardy {\em et al.}\/ through third order in the fields,
we can obtain not only the leading critical behavior
but also the {\em universal}\/ leading corrections to scaling:
these are additive terms of the generic form $\log\log/\log$,
with universal amplitudes that we can compute exactly.
In addition, there are nonuniversal subleading correction-to-scaling
terms of order $1/\log$.
A similar behavior arises in other models with marginally irrelevant
operators, such as the four-dimensional $\varphi^4$ (or Ising) model 
\cite{Brezin_73,Kogon_81},
the three-dimensional tricritical $\varphi^6$ (or spin-1 Ising) model 
\cite{Wegner_73},
and the three-dimensional ferromagnet with strong dipolar interactions
\cite{Larkin_69,Brezin_76}.

A second goal of this paper is to extend the renormalization-group analysis
of Cardy {\em et al.}\/ to obtain the finite-size-scaling (FSS) behavior,
as well as the leading corrections to finite-size scaling,
for the 4-state Potts model in a periodic $L \times L$ box.
Finite-size scaling
\cite{Barber_FSS_review,Cardy_FSS_book,Privman_FSS_book}
plays an important role in the analysis of
Monte Carlo simulations, which of course deal with finite systems
and must thus be extrapolated to the infinite-volume limit.
This extrapolation \cite{Luscher_91,Kim_93,Sokal_extra}
is a very delicate procedure, and the corrections to finite-size scaling
induce systematic errors in the extrapolation.

For $q < 4$ the FSS behavior of the Potts model is well understood.
The critical singularities are rounded-off:
at finite volume, the specific heat (resp.\ the susceptibility) has a peak 
of height $\sim L^{\alpha/\nu}$ (resp.\ $L^{\gamma/\nu}$),
and this peak is shifted from the critical point by an amount
of order $L^{-1/\nu}$. The corrections to this
leading behavior are suppressed by a factor
$L^{-\omega}$, where $\omega$ is a correction-to-scaling exponent.

For $q>4$ the FSS behavior is that appropriate to
a first-order phase transition:
the height of the peak in the specific heat scales as
$\sim L^{d}$, and the shift as $\sim L^{-d}$, where $d$ is the 
dimensionality of the lattice. The corrections are of order 
$L^0$ and $L^{-2d}$, respectively.
This behavior has been proven rigorously for $q \gg1$
by Borgs, Koteck\'y and Miracle-Sol\'e \cite{Borgs_90,Borgs_91,Borgs_92},
and has been confirmed numerically by Billoire {\em et al.}\/ 
\cite{Billoire_93} for $q=20$.

For $q=4$ the FSS behavior is more complicated, but it can be obtained
by a relatively trivial extension of the infinite-volume RG analysis.
Not surprisingly, we find multiplicative logarithmic corrections
to the leading power-law behavior
(some of which have been found previously \cite{Black_Emery});
in addition, we find additive corrections of orders
$\log\log L/\log L$ and $1/\log L$.
In all cases we try to distinguish which corrections are universal
and which are nonuniversal.

We remark that the FSS behavior of the eigenvalues of the transfer 
matrix in an $L \times \infty$ strip is also known 
\cite{Cardy_86,Hamer_88}: there are additive logarithmic corrections 
to scaling due to the marginal dilution field.

The third goal of this paper is to test the predicted finite-size-scaling
behavior for the 4-state Potts model by means of a high-precision
Monte Carlo simulation using a Swendsen-Wang-type algorithm.
We find that our Monte Carlo data are 
consistent with	theory if we include {\em both}\/ the predicted 
multiplicative logarithms {\em and}\/ the additive logarithmic corrections. 
However, it would have been impossible to {\em deduce}\/ the theory from 
these data, which are also consistent with other functional forms. 
In particular, we are unable to observe numerically in an unequivocal way
the correct powers on the multiplicative logarithms
or the correct universal amplitudes on the additive
$\log\log L/\log L$ corrections:
these features are obscured by the presence of the
nonuniversal $1/\log L$ corrections.

\bigskip

The plan of this paper is as follows:
In Section~\ref{sec_fss_infinite} we carry out the RG analysis for
the 4-state Potts model in infinite volume,
with emphasis on the multiplicative logarithms
and on the universal additive logarithmic corrections.
In Section~\ref{sec3} we extend this analysis to determine the
finite-size-scaling behavior and the leading corrections to it.
In Section~\ref{sec_simul} we describe our Monte Carlo simulations.
In Section~\ref{sec_results} we compare our numerical data,
both on and off criticality, to the RG predictions.
In Section~\ref{sec6} we report briefly our data on the
dynamic critical behavior of our Swendsen-Wang-type algorithm.
In the Appendix we show how the Cardy {\em et al.}\/ RG analysis
for $q \approx 4$ can be extended to cubic order in the fields,
yielding the universal third-order term in the RG flow for the dilution field.

%
%
\section{Scaling equations in infinite volume}  \label{sec_fss_infinite}

\subsection{Renormalization-group flow}

It is well known \cite{Nauenberg_80,Cardy_80,Black_Emery} that the 
scaling behavior of the two-dimensional 
4-state Potts model has logarithmic corrections. 
This is due to the presence in this model of a marginal operator,
which is absent in any other two-dimensional Potts model.

Cardy, Nauenberg and Scalapino \cite{Nauenberg_80,Cardy_80} described the 
renormalization-group flow for the $q$-state dilute Potts model by  
using three scaling fields: 
a thermal field $\phi$, an ordering field $h$,
and a dilution field $\psi$.\footnote{
   Of course, in addition to these three fields
   there will be an infinite number of irrelevant scaling fields.
   But these will produce corrections to scaling that are suppressed
   by powers of the length scale, hence are negligible compared to the
   logarithmic corrections arising from the marginal operator $\psi$.
}
The pure Potts model corresponds to a large negative  
value of the dilution field.   
Near criticality, the thermal field is proportional to the temperature
deviation from criticality ($\phi \sim J_c - J$).
The fields $\phi$ and $h$ are relevant, and the field  
$\psi$ is marginal when $q=4$.
The critical point is found at $\phi=h=0$.  
In the larger space $(q,\phi,h,\psi)$,
the point $q=4$, $\phi=h=\psi = 0$ is a multicritical point,
at which a curve of critical points
[$q<4$, $\phi=h=0$, $\psi \sim -(4-q)^{1/2}$]
meets a curve of tricritical points
[$q<4$, $\phi=h=0$, $\psi \sim +(4-q)^{1/2}$].

Cardy {\em et al.}\/  
found that the renormalization-group equations for $q=4$ under an  
infinitesimal change of scale $dl$,
keeping terms through second order in the fields, are 
\begin{subeqnarray} 
{d\psi(l) \over dl} &=& a \psi(l)^2  \\ 
{d\phi(l) \over dl} &=& [y_T + b \psi(l)] \phi(l) \\ 
{d h(l)   \over dl} &=& [y_H + c \psi(l)] h(l)
\label{rg_equations} 
\end{subeqnarray} 
where 
\be 
y_T = {3 \over 2} \, , \quad y_H = {15 \over 8} \, , \quad   
a = {1 \over \pi} \, , \quad b = {3 \over 4 \pi} \, , \quad  
c = {1 \over 16 \pi} \, .  
\ee 
The solution of these equations is: 
\begin{subeqnarray}
\psi(l)  &=&  {\psi(0) \over               1 - a \psi(0)l  } \\
\phi(l)  &=&  {\phi(0) e^{3l/2} \over  (1 - a \psi(0)l)^{3/4}} \\
h(l)     &=&  {h(0)    e^{15l/8}\over  (1 - a \psi(0)l)^{1/16} }
\label{solution_rg_equations} 
\end{subeqnarray}
where $l$ is the logarithm of the length-rescaling factor.
{}From (\ref{solution_rg_equations}b,c) it is clear that the thermal and 
ordering fields are relevant (i.e., they grow exponentially as  
$l\rightarrow\infty$). The dilution field $\psi$ is  
marginally irrelevant when $\psi(0)<0$, in the sense that
$\psi(l)\rightarrow0$ as $l\rightarrow\infty$ but at a subexponential 
rate. This induces multiplicative logarithmic corrections (i.e., powers of $l$) 
in $\phi$ and $h$;  
only when $\psi(0)=0$ do we have a pure power-law (i.e., exponential-of-$l$)
behavior for the two relevant fields.
This latter behavior occurs in the Baxter-Wu 
model \cite{Baxter_Wu}, which is believed to be in
the same universality class as the 
4-state Potts model but does not exhibit any multiplicative logarithmic 
correction. 
We shall henceforth assume without comment that $\psi(0) < 0$.

The equations \reff{rg_equations} are the Poincar\'e normal form
of the RG flow at $q=4$, taking into account terms through
second order in the fields \cite{Cardy_80}.
The analysis of these equations yields the correct leading terms
(power and multiplicative logarithm) for all the
critical observables (correlation length, specific heat, etc.).
However, we should include more terms in order to be able to analyze the 
corrections to scaling to these quantities. Equation (\ref{rg_equations}a) 
should be replaced by  
\be 
{d\psi(l) \over dl} \;\equiv\; F(\psi(l)) \,=\,
    a \psi(l)^2 + a' \psi(l)^3 + \cdots  
\label{rg_equation1_final}
\ee 
where $a'$ is some constant, and the dots $\cdots$ stand for higher powers of
$\psi(l)$.
Note that the ratio  $a'/a^2$ is universal, in the sense that it
cannot be altered by any smooth change of variable;
the coefficients of order $\psi^4$ and higher, by contrast,
can be set to any desired values (e.g.\ zero) by a smooth change of variable
$\psi \to \psi + \alpha_2 \psi^2 + \alpha_3 \psi^3 + \cdots\,$.
The solution of \reff{rg_equation1_final} has the following
asymptotic behavior for large $l$:
\be 
\psi(l) \;=\;  - {1\over al} \left[ 1 + {a' \over a^2} {\log l \over l} + 
             O\!\left( {1\over \l} \right) \right] \; .  
\label{solution_rg_equation1_final}
\ee
Note that the $\log l/l$ correction term here is {\em universal}\/:
it does not depend on $\psi(0)$ or on the parametrization of $\psi$.
By contrast, the $1/l$ corrections are nonuniversal,
as their value depends explicitly on $\psi(0)$.
The terms of order $\psi^4$ and higher in \reff{rg_equation1_final},
even if present, affect only the $1/l$ corrections.

The effect of the term $a' \psi(l)^3$ in \reff{rg_equation1_final}
on the solutions  of   
(\ref{rg_equations}b,c) can be computed easily. Let us define the 
function $f_{A,B}(l;\psi(0))$ to be the ratio 
$\theta(l)/\theta(0)$ when solving the generic equation  
\be 
{d \theta \over dl} = [A + B a \psi(l)] \, \theta(l) \;,  
\label{def_eq_theta} 
\ee
where $\psi(l)$ is the {\em exact}\/ solution of \reff{rg_equation1_final}
with the specified initial value $\psi(0)$.
Namely, the function $f_{A,B}$ is given by the integral 
\begin{subeqnarray}  
f_{A,B}(l;\psi(0))   
                 &=&  \exp \left( \int_0^l [A + B a \psi(l')] \, dl' \right) \\ 
                 &=&  \exp \left( \int_{\psi(0)}^{\psi(l)} 
                      [A + B a\psi'] {d\psi' \over F(\psi') } \right) \;.  
\label{def_fAB} 
\end{subeqnarray}  
The asymptotic behavior of $f_{A,B}$ for large $l$ is
\be 
f_{A,B}(l;\psi(0)) \;=\;  e^{Al} \, \bigl( -a\psi(0) l \bigr)^{-B}
       \left[ 1 + {a' \over a^2} {\log l \over l} + 
               O\!\left( {1 \over l }\right)  \right] ^{B}   \;.
\label{fAB_asymptotic} 
\ee 
Thus, the solutions (\ref{solution_rg_equations}b,c) are replaced by
\begin{subeqnarray}
\phi(l)  &=&  \phi(0) \, f_{{3\over2},{3\over4}}(l;\psi(0))
       \;=\;  {\phi(0) e^{3l/2} \over  (-a \psi(0)l)^{3/4}} 
              \left[ 1 + {a' \over a^2} {\log l \over l}  + 
                     O\!\left( {1 \over \l} \right) \right]^{3/4}  \\
h(l)     &=&  h(0) \, f_{{15\over8},{1\over16}}(l;\psi(0))
       \;=\;  {h(0)    e^{15l/8}\over  (- a \psi(0)l)^{1/16} }
              \left[ 1 + {a' \over a^2} {\log l \over l}  + 
                     O\!\left( {1 \over \l} \right) \right]^{1/16} 
\label{solution_rg_equations_final}
\end{subeqnarray}
One must also include higher powers of $\psi(l)$
inside the square brackets in (\ref{rg_equations}b,c),
but these terms cause only subleading corrections,
multiplying \reff{solution_rg_equations_final}
by factors $1 + O(1/l)$.

The value of $a'$ can be obtained by studying the RG flow in a
neighborhood of the multicritical point $q = q_c = 4$,
and matching the exponents with the exactly known value \cite{Nienhuis_82}
of the next-to-leading thermal exponent as a function of $q$.
The details of this computation are given in 
Appendix~\ref{appendix_rg_flow};
the result is
\be 
a'  \;=\; - {1 \over 2 \pi^2}    \;.
\label{value_aprime} 
\ee
%


\subsection{Correlation length}    \label{sec2.2}

In infinite volume one can define various different correlation lengths:
the most important of these are the exponential correlation length
(= inverse mass gap)
\be
    \xi^{\rm (exp)}_\infty  \;\equiv\;
    \lim\limits_{|x| \to\infty}  {-|x| \over \log G(x)}
 \label{def_xi_exp}
\ee
(where $x$ is taken to infinity along a coordinate axis)
and the second-moment correlation length
\be
    \xi^{\rm (2)}_\infty  \;\equiv\;
    \left( {1 \over 2d} \,
           {\sum\limits_x  |x|^2 \, G(x)
            \over
            \sum\limits_x  G(x) }
    \right) ^{\! 1/2}
    \;,
 \label{def_xi_2}
\ee
where $G(x)$ is the two-point correlation function
and $d$ is the spatial dimension.
(We have appended a subscript $\infty$ to emphasize that these quantities
are defined in infinite volume, in order to distinguish them
from the finite-volume quantities to be introduced in the next section.)
Both of these correlation lengths, as well as other similar ones,
are expected to scale in the same way near the critical point
(but with different prefactors).

Let, therefore, $\xi_\infty$ be any one of these correlation lengths.
It behaves under a change of scale $l$ as 
\be
  \xi_\infty(\phi(0),h(0),\psi(0))  \;=\;
  e^{l} \, \xi_\infty(\phi(l),h(l),\psi(l)) \, .
  \label{scaling_xi_one}
\ee
[Henceforth we shall drop the arguments 0 on the initial values of the fields,
 and simply write $\phi \equiv \phi(0)$, $h \equiv h(0)$,
 $\psi \equiv \psi(0)$.]
Substituting the solutions 
\reff{solution_rg_equations_final},
we get the scaling form
\be
 \xi_\infty(\phi,h,\psi)  \;=\; e^l \,
 \xi_\infty\!\left( \phi f_{{3\over2},{3\over4}}(l;\psi), \, 
                       h f_{{15\over8},{1\over16}}(l;\psi), \,
                    \psi(l)
             \right)  
\label{scaling_xi_fundamental}
\ee
where $\psi(l)$ is given by \reff{solution_rg_equation1_final}.

Let us now define $l^\star \equiv l^\star(\phi,\psi)$
as the solution of the equation
\be
 \phi f_{{3\over2},{3\over4}}(l^\star;\psi) \;=\; \left\{ \begin{array}{ll} 
                 +1 & \quad \hbox{if $\phi>0$} \\
                 -1 & \quad \hbox{if $\phi<0$} \end{array} \right.
\label{l_choice}
\ee
(This solution is unique when $|\phi|$ is sufficiently small.)
The exact form of $l^\star$ as a function of $\phi$ and $\psi$
cannot be obtained in closed form.
However, we can get the following asymptotic expansion for small $\phi$
(and fixed $\psi < 0$):
\begin{eqnarray} 
l^\star(\phi,\psi) &=& - {2 \over 3} \log |\phi|
    \;+\; {1 \over 2}  \log (- \log |\phi| )
    \;+\; {1 \over 2}  \log \!\left( {-2a\psi \over 3} \right)  \nonumber \\ 
  & & \qquad \qquad +\;    
    \left({3 \over 8} - {3 a' \over 4 a^2} \right) 
    {\log(-\log |\phi|) \over -\log |\phi|}   
  \;+\; O\!\left( {1 \over \log|\phi|} \right) \, .
\label{l_star_scaling}
\end{eqnarray} 
Note that the first, second and fourth terms
in \reff{l_star_scaling} are universal,
while the third term and the $O(1/\log |\phi|)$ correction 
are nonuniversal (they depend on $\psi$).
It follows from \reff{l_star_scaling} that for any exponents $A,B$ we can write
\begin{subeqnarray}
F_{A,B}(\phi;\psi) &\equiv& f_{A,B}(l^\star(\phi,\psi);\psi)   \\  
 &=& |\phi|^{-2A/3} \, (-\log |\phi|)^{({A\over 2}-B)} \,
 \left( {-2a\psi \over 3} \right) ^{\! ({A\over 2}-B)} \; \times \nonumber \\
 & & \quad
     \left[ 1 - \left( {3 \over 4} + {3 a' \over 2 a^2} \right)  
                \left( {A\over 2}-B \right)
                { \log(-\log |\phi|)   \over   -\log |\phi| }
             +\,  O\!\left( {1 \over \log |\phi|} \right)  \right]   \;. \qquad
\label{F_AB}
\end{subeqnarray}
Note that the leading power and log are universal, as is the additive
correction proportional to $\log(-\log |\phi|)/\log|\phi|$;
the constant factor and the $1/\log |\phi|$ additive correction
are nonuniversal.

Let us now insert $l = l^\star(\phi,\psi)$ into \reff{scaling_xi_fundamental}:
using \reff{l_star_scaling}/\reff{F_AB},
and restricting for simplicity to zero field ($h=0$),  the result is
\begin{subeqnarray}
 \xi_{\infty}(\phi,0,\psi)  & = &
   F_{1,0}(\phi;\psi) \;
  \xi_{\infty}\!\left( \pm 1,\, 0 ,\, {3 \over 2a \log|\phi|} + \cdots   
              \right) \\[2mm]
   &=& |\phi|^{-2/3} \, \left( {2 a \psi \over 3} \log|\phi| \right)^{1/2} \, 
        \left[ 1 + \left( {3 \over 8} - {3 a' \over 4 a^2} \right)  
               {\log(-\log |\phi|) \over -\log |\phi|} + 
                 O\!\left( {1 \over \log |\phi|}\right) \right] 
  \nonumber  \\ 
  & & \qquad \times \,  
           \xi_{\infty}\!\left( \pm 1,\, 0 \,
                {3 \over 2a\log |\phi|} + \cdots 
              \right) \\[2mm] 
  &\approx& C_\pm \, |\phi|^{-2/3} \, (- \log|\phi|)^{1/2} \,  
        \left[ 1 + \left( {3 \over 8} - {3 a' \over 4 a^2} \right) 
               {\log(-\log |\phi|) \over -\log |\phi|} +
               O\!\left( {1 \over \log |\phi|}\right) \right] 
 \nonumber \\  
 & &  
\label{scaling_xi_phi}
\end{subeqnarray}
where
\be
   C_\pm   \;\equiv\;  \left( {-2a\psi \over 3} \right) ^{1/2} \,
                  \xi_{\infty}(\pm 1, 0,0)
\ee
is a nonuniversal amplitude;
here the $+$ (resp.\ $-$) sign corresponds to the high-temperature
(resp.\ low-temperature) side of criticality.
In the passage from (\ref{scaling_xi_phi}b) to (\ref{scaling_xi_phi}c)
we have assumed that $\xi_{\infty}(\pm1,0,0) \neq 0$ and that
$\xi_{\infty}(\pm1,0,\psi)$ is a {\em smooth}\/ function of $\psi$
near $\psi = 0$:  together these assumptions imply that
$\xi_{\infty}(\pm1, 0, 3/(2a\log |\phi|) + \cdots) =
 {\rm const} + O(1/\log |\phi|)$.
Physically, this amounts to assuming that $\psi$ is a
{\em non-dangerous}\/ irrelevant variable
\cite{Fisher_74,Fisher_83,vEFS}. 
We shall henceforth make this assumption without further comment.
The leading power and multiplicative logarithm in (\ref{scaling_xi_phi}c)
were first obtained in  \cite{Cardy_80};
the universal $\log(-\log |\phi|)/\log |\phi|$ additive correction is new.

If we invert (\ref{scaling_xi_phi}c)  we get
\be
\phi \;\approx\;  \pm C^\prime_\pm \, \xi_\infty^{-3/2} \,
 (\log \xi_\infty )^{3/4}
  \left[ 1 - {3 a' \over 4 a^2} {\log\log\xi_\infty \over \log\xi_\infty}  
         + O\!\left( 1 \over \log \xi_\infty \right) \right] \, ,
\label{phi_vs_xi_infty}
\ee
where
\be 
   C^\prime_\pm  \;\equiv\; (-a \psi)^{3/4} \, \xi_\infty(\pm1,0,0)^{3/2}
\ee
is a nonuniversal amplitude.
Again, the correction of order $\log\log \xi_\infty/\log \xi_\infty$ is 
universal, while the correction of order $1/\log \xi_\infty$ is 
nonuniversal.

\bigskip 
%
%

To study the critical isotherm (with $h >0$) we first define the scale 
$l^{\star\star} \equiv l^{\star\star}(h,\psi)$ to be the solution of 
the equation 
\be 
h f_{{15\over8},{1\over16}}(l^{\star\star};\psi) = 1 
\label{l_choice_h} 
\ee 
This solution is given asymptotically for small $h$ by
\begin{eqnarray} 
l^{\star\star}(h,\psi) &=&  - {8 \over 15} \log h \;+\;
      {1 \over 30} \log(-\log h) \;+\;
      {1 \over 30} \log\left( {-2a\psi \over 15} \right) + \nonumber \\  
 & & \qquad \qquad  
        \left( {1 \over 480} - {a' \over 16 a^2} \right) 
        {\log(-\log h)\over - \log h} \;+\;
        O\!\left( 1 \over \log h \right)  \, .
\label{l_star_star_scaling} 
\end{eqnarray}  
We can also define a function analogous to $F_{A,B}$: 
\begin{subeqnarray} 
G_{A,B}(h;\psi) &\equiv& f_{A,B}(l^{\star\star}(h,\psi);\psi) \\ 
  &=& h^{-8A/15} \, (-\log h)^{{A \over 30} -B} 
      \left( {-8a\psi \over 15} \right)^{{A \over 30} - B} \times  
  \nonumber \\ & &  
      \left[ 1 + \left( {1\over 16} - {15 a' \over 8a^2} \right) 
                 \left( {A \over 30} - B \right) 
                 {\log(-\log h)\over -\log h} + 
                  O\!\left( 1 \over \log h \right) \right] \;. 
\label{G_AB} 
\end{subeqnarray} 
If we now choose $l= l^{\star\star}(h,\psi)$ in \reff{scaling_xi_fundamental},
we get on the critical isotherm ($\phi=0$)
\begin{subeqnarray}  
\xi_\infty(0,h,\psi) &=& G_{1,0}(h;\psi) \;
      \xi_\infty \left( 0,\,
                        1,\,
                        {15 \over 8a}{1 \over \log h} + \cdots
                 \right)  \\  
     &\sim& h^{-8/15} (- \log h)^{1/30} 
\left[ 1 + \left({1\over480} - {a' \over 16a^2} \right) 
         {\log(-\log h) \over -\log h} 
       + O\!\left( 1 \over \log h \right) \right] \;. \nonumber \\  
 & & \slabel{scaling_xi_h2} 
\end{subeqnarray} 

\subsection{The free energy and its derivatives}

The singular piece of the free energy per unit volume behaves under 
a change of scale $l$ as
\be
f_{\rm sing}(\phi(0),h(0),\psi(0)) \;=\;  e^{-2l}
f_{\rm sing}(\phi(l),h(l),\psi(l)) \, .
\label{scaling_fsing_one}
\ee
If we plug the solution \reff{solution_rg_equations_final} of the  
renormalization-group equations \reff{rg_equations} into  
\reff{scaling_fsing_one}, we get
\be
f_{\rm sing}(\phi,h,\psi) \;=\;  e^{-2l}
f_{\rm sing}\!\left( \phi f_{{3\over2},{3\over4}}(l;\psi), \, 
                     h    f_{{15\over8},{1\over16}}(l;\psi), \, 
                     \psi(l)
            \right)
\label{scaling_fsing_fundamental}
\ee
where $\psi(l)$ is given by \reff{solution_rg_equation1_final}.

Let us now choose the scale $l=l^\star(\phi,\psi)$ as in \reff{l_choice};
using \reff{l_star_scaling}/\reff{F_AB}, we get
\be  
 f_{\rm sing}(\phi,h,\psi)  \;=\;
  F_{-2,0}(\phi;\psi) \,
  f_{\rm sing}\!\left( \pm 1,\, h F_{{15\over8},{1\over16}}(\phi;\psi),\, 
              {3 \over 2a \log|\phi|} + \cdots
              \right) \;,  
\label{scaling_fsing_phi_previo} 
\ee 
where the $+$ (resp.\ $-$) sign corresponds to the high-temperature 
(resp.\ low-temperature) side of criticality. 
%
At zero field ($h=0$) this expression can be written as  
\be 
f_{\rm sing}(\phi,0,\psi) \;\approx\;
D_\pm \, |\phi|^{4/3} \,  (-\log|\phi|)^{-1}
       \left[ 1 - \left( {3 \over 4} - {3 a'\over 2 a^2} \right)
                  {\log(-\log |\phi|)\over -\log |\phi|}
        + O\!\left( {1 \over \log |\phi|}\right) \right]
\label{scaling_fsing_phi}
\ee 
where
\be
   D_\pm  \;\equiv\;  \left( {-2a\psi \over 3} \right)^{-1} \,
                      f_{\rm sing}(\pm 1, 0, 0)
\ee
is a nonuniversal amplitude.
The leading power and multiplicative logarithm in \reff{scaling_fsing_phi}
were first obtained in \cite{Nauenberg_80};
the universal $\log(-\log |\phi|)/\log |\phi|$ additive correction is new.

If we differentiate \reff{scaling_fsing_phi_previo}
twice with respect to $\phi$, we get the specific heat.
At zero field ($h=0$) we get   
\begin{eqnarray}
 C_H(\phi,0,\psi) &\sim& |\phi|^{-2/3} 
               (- \log|\phi| )^{-1}  
      \times \nonumber \\[1mm]
  & & \quad  
    \left[ 1 - \left( {3 \over 4} - {3 a'\over 2 a^2} \right) 
         {\log(-\log |\phi|)\over -\log |\phi|}
         + O\!\left( {1 \over \log |\phi|}\right) \right] 
    \;.
\label{scaling_cv_phi} 
\end{eqnarray}
The leading behavior of the specific heat was previously obtained in 
\cite{Nauenberg_80}. 


If we differentiate \reff{scaling_fsing_phi_previo} once with respect 
to the ordering field $h$, we get the magnetization.  
The result at zero ordering field ($h \downarrow 0$) in the low-temperature 
regime ($\phi < 0$) is  
\begin{subeqnarray}  
M(\phi,0,\psi) &\sim& F_{-2,0}(\phi;\psi) \,
                        F_{{15\over8},{1\over16}}(\phi;\psi)
               \;=\;  F_{-{1\over8},{1\over16}}(\phi;\psi)  \\  
   &\sim& 
(-\phi)^{1/12} (- \log(-\phi) )^{-1/8}  \,
 \times \nonumber \\ 
   & & \quad
  \left[ 1 - \left( {3 \over 32} - {3 a'\over 16 a^2} \right) 
         {\log(-\log (-\phi|))\over -\log (-\phi)}
         + O\!\left( {1 \over \log (-\phi)}\right) \right] \;. \quad
\slabel{scaling_m_phi}  
\end{subeqnarray}  
If we differentiate \reff{scaling_fsing_phi_previo} twice with respect to the 
ordering field $h$, we get the susceptibility. The result at $h=0$ is
\begin{subeqnarray}   
\chi(\phi,0,\psi) &\sim& F_{-2,0}(\phi;\psi) \,
                            F_{{15\over8},{1\over16}}(\phi;\psi)^2 
                  \;=\;  F_{{7\over4},{1\over 8}}(\phi;\psi) \\  
 &\sim& 
|\phi|^{-7/6} (-\log|\phi| )^{3/4}  \,
 \times \nonumber \\ 
 & & \qquad
  \left[ 1 - \left( {9 \over 16} - {9 a'\over 8 a^2} \right) 
         {\log(-\log |\phi|)\over -\log |\phi|}
         + O\!\left( {1 \over \log |\phi|}\right) \right] \; . \quad 
\slabel{scaling_chi_phi} 
\end{subeqnarray}  
The leading power and multiplicative logarithms in
\reff{scaling_m_phi}  
were first obtained in \cite{Cardy_80}. 
The result \reff{scaling_chi_phi} did not appear 
in \cite{Cardy_80}, but it can of course be obtained directly from their 
approach.  
The universal $\log(-\log |\phi|)/\log |\phi|$ additive corrections are all new.

Finally, we can write the asymptotic behavior of the specific heat, 
the magnetization,  
and the susceptibility as functions of $\xi_\infty$. Using 
\reff{phi_vs_xi_infty} we get  
\begin{subeqnarray}
C_H   &\sim & { \xi_\infty         \over  (\log \xi_\infty)^{3/2} }  
  \left[ 1 + {3a' \over 2a^2} {\log\log\xi_\infty \over \log\xi_\infty} 
       + O\!\left( 1 \over \log \xi_\infty \right) \right]  
\slabel{scaling_cv_xi_infty} \\[1mm]
M     &\sim & { \xi_\infty^{-1/8}  \over (\log \xi_\infty)^{1/16} } 
  \left[ 1  + {a' \over 16a^2} {\log\log\xi_\infty \over \log\xi_\infty}  
       + O\!\left( 1 \over \log \xi_\infty \right) \right] 
\slabel{scaling_m_xi_infty}  \\[1mm]
\chi  &\sim & { \xi_\infty^{7/4}   \over (\log \xi_\infty)^{1/8} }   
  \left[ 1  + {a' \over 8a^2} {\log\log\xi_\infty \over \log\xi_\infty}  
       + O\!\left( 1 \over \log \xi_\infty \right) \right] 
\, . 
\slabel{scaling_chi_xi_infty} 
\end{subeqnarray}
Note that the (universal) corrections of order
$\log\log\xi_\infty/\log\xi_\infty$ 
are proportional to the ratio $a'/a^2$
and to the exponent of the multiplicative logarithm,
since the other terms of the same order cancel out.  


Finally, we observe that the following two relations hold:
\begin{subeqnarray}
C_H \, \xi_\infty^2  &\sim & \phi^{-2} 
       \,  \left[ 
       1 + O\!\left( {1 \over \log |\phi|} \right) \right] \\[1mm]
M^2 \, \xi_\infty^2  &\sim & \chi
       \,  \left[
       1 + O\!\left( {1 \over \log |\phi|} \right) \right]
\end{subeqnarray}
Note that in these relations there are {\em neither}\/
multiplicative logarithmic corrections {\em nor}\/
additive $\log(-\log |\phi|)/\log |\phi|$ corrections.
It follows that the hyperscaling laws
\begin{subeqnarray}
d\nu &=& 2 - \alpha \\
d\nu &=& \gamma + 2 \beta
\end{subeqnarray}
are satisfied without logarithmic corrections.

\bigskip

The magnetization on the critical isotherm ($\phi=0$, $h>0$) can be calculated 
by first choosing the scale $l=l^{\star\star}(h,\psi)$ as in \reff{l_choice_h}.
Plugging \reff{l_star_star_scaling}--\reff{G_AB} 
into \reff{scaling_fsing_fundamental} we get
\begin{subeqnarray} 
f_{\rm sing}(0,h,\psi) &=& G_{-2,0}(h;\psi) \, 
                f_{\rm sing}\left( 0,\, 1,\, {15 \over 8a \log h} + \cdots
                            \right)  \\ 
   &\sim& 
 h^{16/15} (-\log h)^{-1/15} \, 
 \times \nonumber \\  
   & & \quad  
\left[ 1 - \left( {1 \over 240} - {a'\over 8a^2} \right) 
       {\log(-\log h)\over -\log h} + 
       O\!\left( 1 \over \log h \right) \right] \;.
\end{subeqnarray} 
By differentiating this equation with respect to $h$ we get the critical 
magnetization:
\be
M(0,h,\psi) \sim h^{1/15} (-\log h)^{-1/15} \,
\left[ 1 - \left( {1 \over 240} - {a'\over 8a^2} \right) 
       {\log(-\log h)\over -\log h} + 
       O\!\left( 1 \over \log h \right) \right] 
\, . 
\label{scaling_m_h} 
\ee
The leading power and multiplicative logarithm in this result 
were first obtained also in \cite{Cardy_80}. The universal 
$\log(-\log h)/\log h$ correction is new.

\subsection{Two-point correlators}

The authors of \cite{Cardy_80} also included the following behavior  
for the two-point correlation function at criticality:
\be 
  G(x;\, 0,0,\psi)   \;\sim\;  {1 \over |x|^{1/4} (\log |x|)^{1/8} } 
  \qquad\hbox{as } |x| \to\infty \, . 
  \label{scaling_cor_r} 
\ee
If we assume a natural scaling law of the form 
\be 
  G(x;\, \phi,0,\psi)  \;\approx\;
  {1 \over |x|^{1/4} (\log |x|)^{1/8}} \,
  F_G \!\left( {|x| \over \xi_\infty(\phi,0,\psi)} \right) 
\ee 
where $F_G$ is some scaling function, then
we can deduce equation \reff{scaling_chi_xi_infty} through the integral 
\be 
\chi  \;=\; \int\! d^2x \, G(x)
\, . 
\ee 
Moreover, we can deduce the relation \reff{scaling_m_xi_infty}
by taking $|x| \sim \xi_\infty$
and using the hyperscaling relation
\be
   G(|x| \sim \xi_\infty)   \;\sim\;
   \lim\limits_{|x| \to \infty}  G(x)  \;\equiv\;  M^2  \;.
\ee

\section{Finite-size scaling}  \label{sec3}

In this section we analyze the finite-size-scaling behavior of the
two-dimensional 4-state Potts model on an $L \times L$ lattice
with periodic boundary conditions.

\subsection{Renormalization-group flow in finite volume} 

To obtain the finite-size scaling of this model we only have to adjoin a new
``scaling field'' $L^{-1}$, where $L$ is the linear size of the system.
The fixed-point value of this field obviously is $L^{-1}=0$,
corresponding to the infinite-volume limit. The behavior of this scaling 
field under a change of scale is trivial: to 
(\ref{rg_equations}a--c) we need to adjoin the flow 
\be 
{d L^{-1}(l) \over dl} = L^{-1}(l) \;.  
\label{rg_equations_finiteL} 
\ee

\subsection{Correlation length}

Not all of the correlation lengths employed in infinite volume have
sensible analogues in a fully finite lattice:
for example, the exponential correlation length \reff{def_xi_exp}
makes sense only if the lattice is infinite in at least one direction.
However, the second-moment correlation length \reff{def_xi_2}
does have a sensible extension to finite volume;
this extension is, however, not unique.
One reasonable definition for a periodic lattice of linear size $L$ is
\be 
\xi^{(2)} \;\equiv\; { \left( {\chi \over F}  -1 \right)^{1/2} \over 
                   2\sin (\pi/L) 
                 } \;\, ,  
\ee 
where $\chi$ is the susceptibility
(i.e., the Fourier-transformed two-point correlation function
 at zero momentum)
and $F$ is the corresponding quantity at the smallest nonzero momentum
$(2\pi/L, 0)$:
\begin{eqnarray}
    \chi  & \equiv &   \sum\limits_x  G(x;L)   \\
    F     & \equiv &   \sum\limits_x  e^{2i\pi x_1/L}  G(x;L)
\end{eqnarray}
where $G(x;L)$ is, of course, the two-point correlation function
on the finite lattice.
Another definition could be 
\be 
\xi^{\prime} \;\equiv\;
    \left( {1 \over 2} \,
           {\sum\limits_x  {L^2 \over \pi^2} \sin^2 {\pi x_1 \over L} \, G(x;L)
            \over
            \sum\limits_x  G(x;L) }
    \right) ^{\! 1/2}
    \;=\;
    {L \over 2\pi} \left( 1 - {F \over \chi} \right) ^{\! 1/2}
    \;.
\ee 
Hereafter, $\xi$ will denote any reasonable finite-volume
correlation length.

The generalization of \reff{scaling_xi_one}/\reff{scaling_xi_fundamental} is 
\begin{subeqnarray}
\xi(\phi,h,\psi,L^{-1}) & = &
   e^{l} \, \xi(\phi(l), h(l), \psi(l), L^{-1}(l))     \\[1mm]
 &=& e^{l} \,
         \xi\!\left( \phi f_{{3\over2},{3\over4}}(l;\psi), \,
                      h   f_{{15\over8},{1\over16}}(l;\psi), \,  
                      \psi(l),  
                      {e^l \over L}
            \right) \, ,  
\label{scaling_xi_two} 
\end{subeqnarray}
where $\psi(l)$ is given by \reff{solution_rg_equation1_final}.  
Let us now choose the scale
\be 
  l \;=\; \log L \;,
  \label{l_L}
\ee 
so that \reff{scaling_xi_two} becomes 
\be
\xi(\phi,h,\psi,L^{-1})  \;=\;
L \,
         \xi\!\left( \phi f_{{3\over2},{3\over4}}(\log L;\psi), \,
                      h   f_{{15\over8},{1\over16}}(\log L;\psi), \,  
                      \psi(\log L), \,
          1
   \right)
\, . 
 \label{scaling_xi_fundamental_two} 
\ee

One quantity of interest is the $L$-dependence of the correlation length 
at criticality ($\phi=h=0$). 
{}From \reff{scaling_xi_fundamental_two} and \reff{solution_rg_equation1_final}
it is trivial to see that
\begin{subeqnarray}
\xi(0,0,\psi,L^{-1}) & = &
   L \; \xi\!\left(0,\, 0,\, \psi(\log L) ,\, 1 
                              \right)  \\[1mm]
   & = &
   L \; \left[ x^\star  \,+\,  {A \over \log L}  \,+\,
               B {\log\log L \over (\log L)^2}  \,+\,
               O\!\left( {1 \over (\log L)^2 } \right) \right]  
\label{scaling_xi_L} 
\end{subeqnarray}
where
\begin{subeqnarray}
   x^\star   & \equiv &   \xi(0,0,0,1)  \\[1mm]
   A         & \equiv &   - {1 \over a} \,
                  \left. {\partial \xi \over \partial z}(0,0,z,1)
                  \right| _{z=0}        \\[1mm]
   B         & \equiv &   {a' \over a^2} \, A
\end{subeqnarray}
are all universal;  the corrections of order $1/(\log L)^2$
are nonuniversal.
%
%

We can also obtain a finite-size-scaling law off criticality.
For simplicity let us restrict attention to $h=0$.
{}From \reff{scaling_xi_fundamental_two} we know that at zero field  
\be  
{\xi(\phi,0,\psi,L^{-1}) \over L }  \;=\; 
         \xi\!\left( \phi f_{{3\over2},{3\over4}}(\log L;\psi),\, 0,\,
                      - {1 \over a \log L} + \cdots,\,
          1
   \right)
   \;.
\label{scaling_xi_fundamental_three} 
\ee 
And from \reff{fAB_asymptotic} we know that
\be 
\phi f_{{3\over2},{3\over4}}(\log L;\psi)  \;=\; 
\phi L^{3/2} (-a\psi \log L)^{-3/4}
\left[ 1 + {a'\over a^2} {\log\log L \over \log L} + 
       O\!\left( {1 \over \log L} \right) \right] 
 \label{old5.19}
\ee 
where the $O(1/\log L)$ correction term is nonuniversal.
Therefore, we obtain the FSS law
\be
   {\xi(\phi,0,\psi,L^{-1}) \over L }   \;=\;
   \bar{F}_\xi\bigl( \phi L^{3/2} (\psi \log L)^{-3/4} \bigr)
   \,+\,
   O\!\left( {\log\log L \over \log L} \right)   \;,
 \label{FSS_xi_bare}
\ee
where the function $\bar{F}_\xi$ is universal modulo a nonuniversal
rescaling of its argument.

The disadvantage of this approach is that the corrections to FSS
are of order $\log\log L / \log L$.
A better way to write the FSS law is to consider
the relation between the finite-volume and infinite-volume
correlation lengths at the same $\phi,h,\psi$.
Again restricting for simplicity to $h=0$,
let us insert in \reff{scaling_xi_fundamental_three}
the expression \reff{phi_vs_xi_infty}
for $\phi$ as a function of $\xi_\infty \equiv \xi_\infty(\phi,0,\psi)$.
The first argument on the right-hand side of
\reff{scaling_xi_fundamental_three} can be 
written (for $L, \xi_\infty \gg 1$) as 
%
%
\begin{eqnarray}
&&  \phi \, f_{{3\over2},{3\over4}}(\log L; \psi)  \;=\; \pm E_\pm
       \left( {\xi_\infty \over L} \right)^{\! -3/2}   
       \left( {\log \xi_\infty \over \log L} \right)^{\! 3/4}  \,\times
  \nonumber \\[1mm]
&& \qquad\qquad
\left[ 1 \,+\, {3a' \over 4a^2}
               \left( {\log\log L \over \log L} -
                      {\log\log \xi_\infty \over \log \xi_\infty}
               \right)
\,+\,  {A' \over \log L} \,+\, {B' \over \log \xi_\infty} \,+\, \cdots \right]  
  \qquad
\label{variable} 
\end{eqnarray}
%
where $E_\pm \equiv \xi_\infty(\pm1,0,0)^{3/2}$
%
is a universal amplitude,
the coefficients $A'$ and $B'$ are nonuniversal (they depend on $\psi$),
and the dots denote corrections of order $\log\log L/(\log L)^2$ and
$\log\log \xi_\infty/(\log \xi_\infty)^2$.
If we denote by $\eta \equiv \eta(\psi)$
the coefficient of the nonuniversal $1/l$ correction in 
the formula \reff{fAB_asymptotic} for $f_{A,B}$,
then the values of $A'$ and $B'$ are given by  
\begin{subeqnarray}  
A' &=&  {3 \over 4} \eta(\psi) \\  
B' &=&  - {3 \over 4} \eta(\psi) - {3 \over 4} \log \xi_\infty(\pm1,0,0) 
 - {1 \over a} 
\left. {\partial \over \partial x} \log \xi_\infty(\pm1,0,x) \right|_{x=0}   
\end{subeqnarray} 
%
If we now consider the finite-size-scaling limit
$\phi\to 0$, $\psi$ fixed, $L\to\infty$ with $\xi_\infty/L$ fixed,
we get
\be
\phi \, f_{{3\over2},{3\over4}}(\log L; \psi)  \;=\; \pm E_\pm
       \left( {\xi_\infty \over L} \right)^{\! -3/2}
   \left[ 1 \,+\, {A' + B' + {3 \over 4} \log(\xi_\infty/L)
                   \over \log L}
            \,+\, \cdots \right]
   \;.
\label{variable_bis} 
\ee
Miraculously, the combination $A'+B'$ appearing here
does {\em not}\/ depend on $\psi$. 
Therefore, inserting this into \reff{scaling_xi_fundamental_three}, we obtain
\be 
{\xi(\phi,0,\psi,L^{-1}) \over L }   \;=\;
F_\xi\!\left({ \xi_\infty(\phi,0,\psi) \over L} \right)  \,+\,
\widehat{F}_\xi\!\left( 
           { \xi_\infty(\phi,0,\psi) \over L} \right) {1 \over \log L}  \,+\,
  \cdots  
  \;,
\label{scaling_xi_final} 
\ee  
%
where the functions $F_\xi$ and $\widehat{F}_\xi$ are {\em universal}\/
(they do not contain any piece depending on $\psi$).
It is quite remarkable that not only the leading finite-size-scaling
function, but also the leading correction to it, is universal
(in amplitude as well as in shape).
This is a special feature of the logarithmic corrections to scaling induced by
a marginally irrelevant operator, and is not observed in the more common
context of power-law corrections to scaling induced by irrelevant operators.

The equation \reff{scaling_xi_final} can be formally inverted, yielding
\be
{\xi_\infty(\phi,0,\psi) \over L }   \;=\;
G_\xi\!\left({ \xi(\phi,0,\psi,L^{-1}) \over L} \right)  \,+\,
\widehat{G}_\xi\!\left( 
          { \xi(\phi,0,\psi,L^{-1}) \over L} \right) {1 \over \log L} 
 \,+\, \cdots 
\label{scaling_xi_infty_final} 
\, .
\ee
%
Again both functions  $G_\xi$ and $\widehat{G}_\xi$ are {\em universal}\/.

\subsection{The free energy and its derivatives}

The finite-volume generalization of \reff{scaling_fsing_fundamental} is
\be
f_{\rm sing}(\phi,h,\psi,L^{-1}) = e^{-2l}
f_{\rm sing}\!\left( \phi f_{{3\over2},{3\over4}}(l;\psi),\,  
                      h   f_{{15\over8},{1\over16}}(l;\psi),\,
                      \psi(l),\,
                   {e^l \over L}
            \right) \,, 
\label{scaling_fsing_two}
\ee
where $\psi(l)$ is given by \reff{solution_rg_equation1_final}. 
As in \reff{l_L} we choose $l = \log L$, and get
\be 
f_{\rm sing}(\phi,h,\psi,L^{-1}) = L^{-2} 
f_{\rm sing}\!\left( \phi f_{{3\over2},{3\over4}}(\log L;\psi),\,   
                     h   f_{{15\over8},{1\over16}}(\log L;\psi),\,  
                     -{1 \over a\log L}+\cdots,\,  
                     1
   \right) \, . 
 \label{scaling_fsing_fundamental_two}
\ee  
The finite-size behavior near criticality of the specific heat, magnetization, 
and susceptibility can be obtained by performing the appropriate derivatives
with respect to $\phi$ or $h$.
For simplicity let us restrict attention to zero field,
where we get\footnote{
   We refrain from writing also
   \begin{eqnarray*}
M(\phi,0,\psi,L^{-1})  &=&  {L^{-1/8} \over (-a \psi \log L)^{1/16} }   
\left[ 1 + {a'\over 16a^2} {\log\log L\over \log L} +
     {k_M \over \log L} + \cdots \right]
                       \times \nonumber \\
        & & \qquad {\partial f_{\rm sing} \over \partial y} \!
        \left. \left( \phi f_{{3\over2},{3\over4}}(\log L;\psi),\, y,\,  
                      -{1\over a\log L} +\cdots,\, 1 \right)
        \right|_{y=0}
   \end{eqnarray*}
   because the finite-volume magnetization of course {\em vanishes}\/
   at $h=0$, for all $\phi$ (by virtue of the Potts symmetry).
   This scaling is indeed valid, but the prefactor
   $(\partial f_{\rm sing} / \partial h)(\phi,h,\psi,1)$
   {\em vanishes}\/ at $h=0$.

   The scaling $L^{-1/8} (\log L)^{-1/16}$ at criticality
   applies not to the usual magnetization $M \equiv L^{-2} \< {\cal M} \>$,
   but rather to the ``absolute magnetization''
   $\widetilde{M} \equiv L^{-2} \< |{\cal M}| \>$.
   These two quantities are essentially identical in the
   low-temperature phase ($h \downarrow 0$ at {\em fixed}\/ $\phi < 0$),
   but are very different in the critical regime.
   Unfortunately, much of the literature (especially numerical work)
   has sloughed over the distinction between $M$ and $\widetilde{M}$.
}
\begin{eqnarray} 
C_H(\phi,0,\psi,L^{-1}) &=& {L \over (-a \psi \log L)^{3/2} }  
\left[ 1 + {3a'\over 2a^2} {\log\log L\over \log L} + 
     {k_{C_H} \over \log L} + \cdots \right] 
                       \times \nonumber \\ 
        & & \qquad {\partial^2 f_{\rm sing} \over \partial x^2} \! 
        \left. \left( x,\, 0,\,  -{1\over a\log L}+\cdots ,\, 1 \right) 
        \right|_{x = \phi f_{{3\over2},{3\over4}}(\log L;\psi)} 
\label{scaling_cv_general_L} \\[5mm] 
\chi(\phi,0,\psi,L^{-1})  &=&  {L^{7/4} \over (-a \psi \log L)^{1/8} }
\left[ 1 + {a'\over 8a^2} {\log\log L\over \log L} +
     {k_\chi \over \log L}  + \cdots \right]
                       \times \nonumber \\
        & & \qquad {\partial^2 f_{\rm sing} \over \partial y^2} \!
        \left. \left( \phi f_{{3\over2},{3\over4}}(\log L;\psi),\, y,\,  
                      -{1\over a\log L}+\cdots,\, 1  \right)
        \right|_{y=0}
   \qquad
\label{scaling_chi_general_L} 
\end{eqnarray} 
%
%
For future convenience, we have denoted by $k_{\cal O}$
the coefficient of the nonuniversal $1/\log L$ correction term
within the square brackets,
which arises from the nonuniversal $1/l$ term in \reff{fAB_asymptotic}.
In addition to the two correction terms within the square brackets,
we have (through order $1/\log L$) three other corrections arising from
the factors involving $f_{\rm sing}$:
a {\em universal}\/ $1/\log L$ term arising from the derivative
of the relevant scaling function with respect to $\psi$ at $\psi=0$;
and a {\em universal}\/ $\log\log L/\log L$ term
plus a {\em nonuniversal}\/ $1/\log L$ term,
both arising from the fact
that the first argument of $f_{\rm sing}$ is not exactly
proportional to $\phi L^{3/2} (\log L)^{-3/4}$, but is rather
given by \reff{old5.19}.
The simplest case is {\em at}\/ criticality ($\phi=0$):
then the second and third terms vanish, and the first one gets amalgamated
with the nonuniversal terms $k_{\cal O}/\log L$.
We thus get
\begin{eqnarray} 
C_H(0,0,\psi,L^{-1}) &\sim& {L \over (\log L)^{3/2}} \, 
    \left[ 1 + {3a'\over 2a^2} {\log\log L\over \log L} +
     {k'_{C_H} \over \log L} + \cdots \right] 
 \label{scaling_cv_L} \\[1mm]
\chi(0,0,\psi,L^{-1}) &\sim& {L^{7/4} \over (\log L)^{1/8}} \, 
    \left[ 1 + {a'\over 8a^2} {\log\log L\over \log L} +
     {k'_\chi \over \log L} + \cdots \right] 
 \label{scaling_chi_L} 
\end{eqnarray} 
where $k'_{C_H}$ and $k'_\chi$ are nonuniversal.

In conclusion, all the observables which are computed through
derivatives of the free energy have at criticality
a multiplicative logarithmic piece, proportional to some power of $\log L$.
They also have additive corrections of order 
$\log\log L/\log L$, the coefficient of which is $-a'/a^2$ times the
exponent of the multiplicative logarithm.
These corrections arise together from the fact that with each 
derivative we gain a factor $f_{A,B}(\log L;\psi)$.  
On the other hand, the $1/\log L$ corrections in 
\reff{scaling_cv_L}/\reff{scaling_chi_L} are nonuniversal 
because the corresponding terms
in $f_{A,B}(\log L;\psi)$ are nonuniversal. 
Both the multiplicative logarithm and the additive $\log\log L/\log L$ term
are absent in the correlation length \reff{scaling_xi_L}.
The leading power and multiplicative logarithm in
\reff{scaling_cv_L} were obtained previously in \cite{Black_Emery}. 
It is noteworthy to remark that in some of
the literature it was (wrongly!) assumed that the leading term for the 
susceptibility had no such logarithmic corrections 
\cite{Wiseman_Domany,Salas_Sokal_LAT95,Salas_Sokal_AT}.  

Off criticality, we get FSS laws
\begin{eqnarray}
C_H(\phi,0,\psi,L^{-1}) &\sim& {L \over (\log L)^{3/2}}
   \left[ \bar{F}_{C_H}\bigl( \phi L^{3/2} (\psi \log L)^{-3/4} \bigr)
   \,+\,
   O\!\left( {\log\log L \over \log L} \right) \right] \nonumber \\
 & &
 \label{scaling_bare__cv_L} \\[1mm]
\chi(\phi,0,\psi,L^{-1}) &\sim& {L^{7/4} \over (\log L)^{1/8}}
   \left[ \bar{F}_{\chi}\bigl( \phi L^{3/2} (\psi \log L)^{-3/4} \bigr)
   \,+\,
   O\!\left( {\log\log L \over \log L} \right) \right] \nonumber \\
 & &
 \label{scaling_bare_chi_L}
\end{eqnarray}
The disadvantage of \reff{scaling_bare__cv_L}/\reff{scaling_bare_chi_L}
is that the corrections to FSS
are of order $\log\log L / \log L$.
A better way to write the FSS law is to use $\xi_\infty/L$ as the
argument of the scaling function.
Using the relation \reff{variable_bis}, we get
\begin{eqnarray} 
C_H(\phi,0,\psi,L^{-1}) &\sim& {L \over (\log L)^{3/2}} 
\left[ 1 + {3a'\over 2a^2} {\log\log L\over \log L}  + 
{k_{C_H} \over \log L} + \cdots \right]  \times \nonumber \\
 & & \quad
\left[ F_{C_H}\!\left({ \xi_\infty(\phi,0,\psi) \over L} \right)  \,+\,
\widehat{F}_{C_H}\!\left({ \xi_\infty(\phi,0,\psi) \over L} \right) 
     {1 \over \log L} + \cdots \right] \nonumber \\  
  &  & \\ 
\chi(\phi,0,\psi,L^{-1}) &\sim& {L^{7/4} \over (\log L)^{1/8}} 
\left[ 1 + {a'\over 8a^2} {\log\log L\over \log L}  + 
{k_\chi \over \log L} + \cdots \right]  \times \nonumber \\
 & & \quad
\left[ F_{\chi}\!\left({ \xi_\infty(\phi,0,\psi) \over L} \right)  \,+\,
\widehat{F}_{\chi}\!\left({ \xi_\infty(\phi,0,\psi) \over L} \right)
     {1 \over \log L} + \cdots \right] 
\end{eqnarray} 
where the scaling functions $F_{\cal O}$ and $\widehat{F}_{\cal O}$
are {\em universal}\/ and can be
expressed in terms of derivatives of $f_{\rm sing}$.
Thus, the nonuniversal terms $k_{\cal O}/\log L$
appear only in the prefactor in square brackets.
There is, of course, also a nonuniversal prefactor multiplying everything.

Finally, we can re-express $\xi_\infty/L$ as a function of 
$\xi(\phi,L)/L \equiv \xi(\phi,0,\psi,L^{-1})/L$, 
using \reff{scaling_xi_infty_final},
and get 
\begin{eqnarray}
C_H(\phi,0,\psi,L^{-1}) &\sim& {L \over (\log L)^{3/2}} 
\left[ 1 + {3a'\over 2a^2} {\log\log L\over \log L} +  
     {k_{C_H} \over \log L} + \cdots \right] \times \nonumber \\
 & & \quad
\left[ G_{C_H}\!\left({ \xi(\phi,L) \over L} \right)  \,+\,
\widehat{G}_{C_H}\!\left({ \xi(\phi,L) \over L} \right)
     {1 \over \log L} + \cdots \right]  
\label{scaling_cv_vs_xi} 
\\[2mm]
\chi(\phi,0,\psi,L^{-1}) &\sim& {L^{7/4} \over (\log L)^{1/8}} 
\left[ 1 + {a'\over 8a^2} {\log\log L\over \log L} +
     {k_\chi \over \log L} + \cdots \right] \times \nonumber \\
 & & \quad
\left[ G_{\chi}\!\left({ \xi(\phi,L) \over L} \right)  \,+\,
\widehat{G}_{\chi}\!\left({ \xi(\phi,L) \over L} \right)
     {1 \over \log L} + \cdots \right]  
\label{scaling_chi_vs_xi}  
\end{eqnarray}
%
The scaling functions $G_{\cal O}$ and $\widehat{G}_{\cal O}$ are universal.


If we study the ratio $C_H(\phi,0,\psi,L^{-1})/C_H(0,0,\psi,L^{-1})$, we 
can see easily from \reff{scaling_cv_vs_xi} that
the prefactor in square brackets cancels out, and we have
\be
{C_H(\phi,0,\psi,L^{-1}) \over C_H(0,0,\psi,L^{-1}) }  \;\approx\;
G_{C_H}\left( {\xi(\phi,L) \over L} \right) + 
\widehat{G}_{C_H}^{(1)} 
     \left( {\xi(\phi,L) \over L} \right) {1 \over \log L} + \cdots \;,  
\ee
where both scaling functions are {\em universal}\/.
Therefore, both the universal $\log\log L/\log L$ corrections and
the nonuniversal $1/\log L$ corrections 
to the specific heat $C_H(\phi,0,\psi,L^{-1})$ come from its value {\em at}\/
criticality. Similar reasoning applies to the susceptibility.

%
\section{Description of the simulations} \label{sec_simul} 

\subsection{The Monte Carlo algorithm} \label{sec_algorithm}

The Monte Carlo (MC) algorithm
we used in our simulations was actually an algorithm
\cite{Salas_Sokal_LAT95,Salas_Sokal_AT}
to simulate the Ashkin-Teller (AT) model \cite{Ashkin_Teller,Baxter}.  
This model is  
a generalization of the Ising model to a four-state model, and it includes 
as a particular case the 4-state Potts model. The general AT model 
assigns to each lattice site $x$ two Ising spins $\sigma_x = \pm 1$  
and $\tau_x = \pm 1$, and they interact through the Hamiltonian
\be
\label{AT_hamiltonian}
H_{\rm AT} \;=\; -J \sum_{\< xy \>} \sigma_x \sigma_y
             -J'\sum_{\< xy \>} \tau_x \tau_y
             -K \sum_{\< xy \>} \sigma_x  \tau_x \sigma_y \tau_y \; ,
\ee
where the sums run over nearest-neighbor pairs $\< xy \>$. The line 
$J=J'=K$ is the 4-state Potts model with $\beta = 4 J$:
\be 
\label{Potts_hamiltonian} 
H_{\rm Potts} \;=\; -J \sum_{\< xy \>} (\sigma_x \sigma_y + 
\tau_x \tau_y + \sigma_x  \tau_x \sigma_y \tau_y) \;=\; 
-4J \sum_{\< xy \>} \delta_{\sigma_x,\sigma_y} 
\delta_{\tau_x,\tau_y} \,+\, {\rm const}  \;.  
\ee 
The critical point of the square-lattice 4-state Potts model
lies at $J=J'=K= {1 \over 4} \log 3$.  
The plane $K=0$ of the general AT model corresponds to two 
non-interacting Ising models with nearest-neighbor constants 
$J$ and $J'$, respectively.

Wiseman and Domany \cite{Wiseman_Domany}   
proposed an  algorithm of  Swendsen-Wang (SW) type for the general AT 
model. This algorithm (called the ``direct algorithm'' in 
\cite{Salas_Sokal_LAT95,Salas_Sokal_AT}) reduces to the standard 
SW algorithm \cite{Swendsen_87} at the Ising and 4-state Potts subspaces.  
In \cite{Salas_Sokal_LAT95,Salas_Sokal_AT} we proposed an embedding 
variant of their algorithm;  it reduces to the standard SW algorithm 
at the Ising plane, but not at the 4-state Potts line. However, we gave
numerical evidence that the two algorithms lie in the same dynamic universality
class.
In particular, we found that at criticality
\be
{\tau_{{\rm int}, {\cal E}}^{\rm direct} \over
 \tau_{{\rm int}, {\cal E}}^{\rm embedd} } \;=\; 1.516 \pm 0.035
 \;,
\label{tau_ratio}
\ee
where $\tau_{{\rm int}, {\cal E}}$ denotes the integrated  
autocorrelation time of the energy
(this is roughly the slowest mode in SW-type dynamics).
However, because the embedding algorithm requires 1.9 times as much CPU time
per iteration as the direct algorithm (since in the AT formulation there
are twice as many spin variables as in the Potts formulation),
our algorithm is about 25\% less efficient than the standard SW algorithm.

Let us review briefly our embedding algorithm (more details can be 
found in \cite{Salas_Sokal_AT}). First, consider 
the Boltzmann weight of a given bond $\<xy\>$, {\em conditional on the 
$\{\tau\}$ configuration} (i.e., the $\tau$ spins are kept fixed): it is 
\be
\label{Boltzmann_weight_sigma}
W_{\rm bond}(\sigma_x,\sigma_y;\tau_x,\tau_y) \;=\;
e^{-2J(1+\tau_x\tau_y)} +
\left[ 1 - e^{-2J(1+\tau_x\tau_y)} \right]
\delta_{\sigma_x,\sigma_y} \; .
\ee
We can simulate this system of $\sigma$ spins using a
standard SW algorithm. The effective nearest-neighbor coupling
\be
\label{effective_coupling}
J_{xy}^{\rm eff}  \;=\;  J (1 +  \tau_x \tau_y) 
\ee
is no longer translation-invariant,
but this does not matter. The key point is that
the effective coupling is always {\em ferromagnetic}. 
An exactly analogous argument applies to the $\{\tau\}$ spins when the
$\{\sigma\}$ spins are held fixed.

The embedding algorithm for the 4-state Potts model has therefore two parts:

\medskip

{\bf Step 1:  Update of $\{\sigma\}$ spins.}
Given the $\{\tau\}$ configuration (which we hold fixed),
we perform a standard SW iteration on the $\sigma$ spins.
The probability $p_{xy}$ arising in the SW algorithm takes the value
$p_{xy} = 1 - \exp[-2J(1+\tau_x\tau_y)]$.

\medskip

{\bf Step 2:  Update of $\{\tau\}$ spins.}
Given the $\{\sigma\}$ configuration (which we hold fixed),
we perform a standard SW iteration on the $\tau$ spins.
The probability $p_{xy}$ arising in the SW algorithm takes the value
$p_{xy} = 1 - \exp[-2J(1+\sigma_x\sigma_y)]$.

\medskip

One iteration of the embedding algorithm consists, by definition,
of a single application of Step 1 followed by a single application of Step 2.

\subsection{Observables to be measured} \label{sec_obs}

Let us begin by defining some basic observables.
The observables of interest involving only the $\sigma$ spins are
\begin{eqnarray}
\label{Msigma}
{\cal M}_\sigma    & \equiv & \sum_x \sigma_x  \\[2mm]
\label{Esigma}
{\cal E}_\sigma    & \equiv & \sum_{\<xy\>} \sigma_x \sigma_y \\[2mm]
\label{Fsigma}
{\cal F}_\sigma    & \equiv & {1 \over 2} \left[
                       \left| \sum_x \sigma_x e^{2i\pi x_1 / L}
                       \right|^2 +
                       \left| \sum_x \sigma_x e^{2i\pi x_2 / L}
                       \right|^2
                       \right]
\end{eqnarray}
where $L$ is the linear size of the system
(we always use periodic boundary conditions)
and $(x_1,x_2)$ are the
Cartesian coordinates of the point $x$. The observable ${\cal F}_\sigma$
can be also seen as the square of the Fourier transform of $\sigma$
at the smallest allowed non-zero momenta
[i.e., $(\pm 2 \pi/L,0)$ and $(0,\pm 2 \pi/L)$ for the square lattice];
it is normalized to be comparable to its zero-momentum analogue
${\cal M}_\sigma^2$.
We define analogous observables for the $\tau$ spins and for the
composite operator $\sigma\tau$.

At the 4-state Potts line we have a symmetry under permutations of 
$(\sigma,\tau,\sigma\tau)$. Thus, the natural choice of observables 
are those invariant under this symmetry. We have measured the 
following observables:
\begin{eqnarray}
{\cal M}^2 &\equiv& \smfrac{1}{3} \left( {\cal M}^2_\sigma + {\cal M}^2_\tau +
                                  {\cal M}^2_{\sigma\tau} \right) \\
{\cal E}   &\equiv& \smfrac{1}{3} \left( {\cal E}_\sigma + {\cal E}_\tau +
                                  {\cal E}_{\sigma\tau} \right) \\
{\cal F}   &\equiv& \smfrac{1}{3} \left( {\cal F}_\sigma + {\cal F}_\tau +
                                  {\cal F}_{\sigma\tau} \right)
\end{eqnarray}
These observables coincide with the usual ones for the 4-state Potts model
up to some multiplicative constants.
We then define the magnetic susceptibility
\be
\chi \;=\; {1 \over V} \< {\cal M}^2 \>   \;,
\ee
the two-point correlation at the smallest nonzero momentum
\be
F  \;=\;  {1 \over V} \< {\cal F} \>   \;,
\ee
the second-moment correlation length
\be
\xi^{(2)} \;=\; { \left(\! {\chi \over F} - 1 \!\right)^{\! 1/2}
         \over   2 \sin {\pi \over L} }
\; , 
\ee
the energy density (per bond) 
\be
E \;=\; {1 \over 2V} \< {\cal E} \>   \;,
\ee
and the specific heat
\be
C_H \;=\; {1 \over 2V} \left( \< {\cal E}^2 \> - \< {\cal E} \>^2 \right)
 \;.
\ee
In all these formulae, $V=L^2$ is the number of lattice sites
and $2V$ is the number of bonds
(we have a square lattice with periodic boundary conditions).

Finally, let us define the quantities associated with the Monte Carlo
dynamics.
Given an observable ${\cal O}$, we define the corresponding 
unnormalized autocorrelation function as 
\be
C_{\cal OO}(t) \;=\; \< {\cal O}_s {\cal O}_{s+t} \> - \< {\cal O} \>^2
\;,
\ee
where all the expectation values $\< \cdot \>$ are taken in equilibrium, and
$t$ is the ``time'' in units of MC steps. 
The associated normalized autocorrelation function is
\be
\rho_{\cal OO}(t) \;=\; {C_{\cal OO}(t) \over  C_{\cal OO}(0) } 
\;.
\ee
The integrated autocorrelation time for the observable ${\cal O}$ is
defined as
\be 
\tau_{{\rm int},{\cal O}} = {1 \over 2} \sum_{t=-\infty}^{\infty}
                            \rho_{ \cal O O }(t) \; .  
\label{def_tau_int} 
\ee
The integrated autocorrelation time controls the statistical error in 
MC estimates of the mean $\< {\cal O} \>$.
Given a time series of measurements
$\{ {\cal O}_1, {\cal O}_2, \ldots, {\cal O}_n \}$
(in equilibrium),
the sample mean
\be
\overline{{\cal O}}  \;\equiv\;  {1 \over n} \sum_{t=1}^n {\cal O}_t
\ee
constitutes an unbiased estimator of $\< {\cal O} \>$,
and its variance (when $n\gg \tau_{{\rm int},{\cal O}}$) is
\be 
{\rm var}(\overline{{\cal O}}) = {1 \over n}\ 2 \tau_{{\rm int},{\cal O}}  
         C_{\cal OO}(0) 
\; . 
\ee 
Thus, the variance of $\overline{\cal O}$ is 
$2 \tau_{{\rm int},{\cal O}}$ larger than it would be if the measurements 
were uncorrelated.  We can numerically obtain reliable estimates of 
both $\tau_{{\rm int},{\cal O}}$ and its error bar (from the autocorrelation 
function) by using a self-consistent truncation procedure
\cite[Appendix C]{Madras_Sokal}.  
We have used a window of width $6 \tau_{{\rm int},{\cal O}}$, which is 
sufficient whenever the autocorrelation function decays roughly 
exponentially. This almost-exponential decay has been confirmed numerically in 
\cite{Salas_Sokal_AT}.

   To compute the specific-heat error-bar we used the following procedure: 
first we computed the mean energy $\< {\cal E} \>$, and then considered 
the observable ${\cal O} \equiv ({\cal E} - \< {\cal E} \>)^2$ using 
using the general procedure described in this section.

\subsection{Summary of the simulations} \label{sec_summary} 

We have simulated the 4-state Potts model on an $L \times L$ square lattice
with periodic boundary conditions,
using the ``embedding'' algorithm described in Section~\ref{sec_algorithm}.
We performed the simulation at 194 pairs $(J,L)$.
The values of the coupling constant $J$ range from 0.261 to the critical point 
$J_c = {1 \over 4} \log 3 \approx 0.274653$. 
The lattice sizes $L$  range from 16 to 1024 at the critical point,
and from 32 to 512 off criticality.
In all cases we have started our simulations with a random configuration, 
and we have discarded the first $10^5$ iterations to allow the system to reach 
equilibrium. This discard interval is more than sufficient:
in the worst case ($L=1024$ at $J_c$) 
it is roughly equal to $190\tau_{{\rm int},{\cal E}}$ (or 
$160\tau_{{\rm exp},{\cal E}}$ \cite{Salas_Sokal_AT}),
and in all other cases it is at least $300\tau_{{\rm int},{\cal E}}$.

The length of the runs ranges from $9\times 10^5$ to $10^7$ iterations. For 
$L=512$ this run length corresponds to $10^4$ times $\tau_{{\rm int},{\cal E}}$;
for $L=256$ it is at least $2 \times 10^4 \tau_{{\rm int},{\cal E}}$;
for $L=128$, at least $3 \times 10^4 \tau_{{\rm int},{\cal E}}$;
and for $L=64$, at least $4 \times 10^4 \tau_{{\rm int},{\cal E}}$.
These run lengths are more than sufficient to get a fairly good determination
of the dynamic quantities, and to get high-precision data for the static 
quantities.  
Unfortunately,
for $L=1024$ we were able to achieve only $1500 \tau_{{\rm int},{\cal E}}$,
as the autocorrelation time is quite large \cite{Salas_Sokal_AT}.

The data at the critical point were employed already in our previous
work \cite{Salas_Sokal_LAT95,Salas_Sokal_AT}
to extract the dynamic critical behavior of our SW-type algorithm.
Here we have improved the statistics at $L=64,128,256$;
the revised data at criticality are in displayed in 
Table~\ref{table_mc_data_tc}.  
The whole set of data, including the 187 runs off criticality,
can be obtained from the authors.

The CPU time required by our program is approximately $10 L^2$ $\mu$s/iteration 
on an IBM RS-6000/370. The total CPU time used in this project was 
approximately 8.5 years on this machine. The simulations were mainly run on 
the CAPC cluster at New York University, the IBM SP2 cluster at the Cornell 
Theory Center, and the DEC Alpha cluster at the
Pittsburgh Supercomputing Center.

\section{Numerical Results} \label{sec_results}

\subsection{Finite-size scaling at criticality} \label{sec_res_crit}

In this subsection we are going to test the 
FSS predictions 
\reff{scaling_xi_L}/\reff{scaling_cv_L}/\reff{scaling_chi_L} 
for the 4-state Potts model {\em at criticality}\/.
In \cite{Salas_Sokal_LAT95,Salas_Sokal_AT} we presented some preliminary  
results. However, we have now made significant extensions of these runs,
in some cases doubling the statistics.
More importantly, we now have a better {\em theoretical}\/ knowledge of the 
FSS behavior of this model, which includes both multiplicative and
additive logarithmic corrections.

For each quantity ${\cal O}$, we shall carry out fits to  
several different Ans\"atze using the standard weighted least-squares method.
As a precaution against corrections to scaling,
we impose a lower cutoff $L \ge L_{min}$
on the data points admitted in the fit,
and we study systematically the effects of varying $L_{min}$ on both 
the estimated parameters and the $\chi^2$.
In general, our preferred fit corresponds to the smallest $L_{min}$
for which the goodness of fit is reasonable
(e.g., the confidence level\footnote{
   ``Confidence level'' is the probability that $\chi^2$ would
   exceed the observed value, assuming that the underlying statistical
   model is correct.  An unusually low confidence level
   (e.g., less than 5\%) thus suggests that the underlying statistical model
   is {\em incorrect}\/ --- the most likely cause of which would be
   corrections to scaling.
}
is $\gtapprox$ 10--20\%)
and for which subsequent increases in $L_{min}$ do not cause the
$\chi^2$ to drop vastly more than one unit per degree of freedom.

\subsubsection{Second-moment correlation length}

The quantity $\xi^{(2)}/L$ is expected to approach a 
constant $x^\star$ as $L\to\infty$, with additive $O(1/\log L)$
corrections \reff{scaling_xi_L}. The constant $x^\star$ can be 
in principle computed via conformal field theory, but to our 
knowledge this calculation has not yet been done.

A fit of this ratio to a constant is reasonably good only for 
$L_{min}=128$:
\be 
{\xi^{(2)}\over L} = x^\star = 1.0022 \pm 0.0023 
\ee 
with $\chi^2=2.38$ (3 DF, level = 50\%). However, if we take into 
account the $1/\log L$ corrections we get a good fit already for $L_{min}=16$:
\be 
{\xi^{(2)}\over L} = (1.0221 \pm 0.0061) - {0.1077 \pm 0.0239 \over \log L} 
\ee 
with $\chi^2=2.32$ (5 DF, level = 80\%).
%
%
On the other hand, an equally good fit can 
be obtained using a power-law correction $L^{-\Delta}$ with 
$0 < \Delta \ltapprox 0.5$;
this is to be expected, as a logarithm can be well mimicked by a small power.
We can conclude conservatively that
\be
   x^\star  \;=\;  1.02 \pm 0.03  \;.
\ee

\subsubsection{Susceptibility}

The expected behavior of the critical susceptibility at 
finite $L$ is given by \reff{scaling_chi_L}. Thus, in 
addition to the standard $L^{\gamma/\nu}$ term with $\gamma/\nu=7/4$, we have 
a multiplicative logarithmic correction $(\log L)^{-1/8}$, and also 
additive logarithmic corrections of orders $\log\log L/\log L$ and $1/\log L$. 
All these features make the 
accurate numerical estimation of the exponent $\gamma/\nu$ quite difficult.

If we fit the susceptibility to the naive power law 
$A L^p$, we get a reasonably good fit for $L_{min}=16$:
\be 
{\gamma \over \nu} = 1.744 \pm 0.001 
 \label{eq5.3}
\ee
with $\chi^2=2.21$ (5 DF, level = 82\%). The difference 
from the exact result 1.75 is not large,
but it is six standard deviations and thus strongly
statistically significant. The fact that the estimate \reff{eq5.3}
estimate is {\em smaller}\/ than the exact value is consistent with 
the existence of a multiplicative logarithmic correction raised to a 
{\em negative}\/ power. However, 
without the theoretical knowledge of a multiplicative 
logarithmic correction, we could equally well conclude that this small
discrepancy is due to additive corrections to scaling 
\cite{Salas_Sokal_AT}.

If we try to fit the quantity $\chi / L^{7/4}$ to an 
arbitrary power of $\log L$ [i.e., $\chi / L^{7/4} = A (\log L)^p$] 
we obtain a reasonable fit already for $L_{min}=16$:
\be 
p = -0.0236\pm 0.0040  
\label{res_p_improved} 
\ee 
with $\chi^2=3.64$ (5 DF, level = 60\%). This result differs drastically  
from the theoretical prediction $p=-1/8=-0.125$.
Alternatively, we can try to fit
$\chi (\log L)^{1/8}$ to a power-law $A L^{\gamma/\nu}$; the result for 
$L_{min}=128$ is 
\be 
{\gamma \over \nu} = 1.765 \pm 0.003 
\label{res_gamma_over_nu_improved} 
\ee 
with $\chi^2=1.83$ (2 DF, level = 40\%). Once again this differs by five 
standard deviations from the exact value 1.75.

%
Looking at the FSS prediction \reff{scaling_chi_L}, it is plausible to 
think that the  
reason why \reff{res_p_improved}/\reff{res_gamma_over_nu_improved} 
differ so radically from the leading-order theoretical prediction 
$L^{7/4} (\log L)^{-1/8}$ are the large corrections to scaling. To test this 
idea, we have fitted $\chi/[L^{7/4} (\log L)^{-1/8}]$ to 
$A + B \log\log L/\log L$. The fit is reasonable for $L_{min}=64$:
\be 
{ \chi \over L^{7/4} (\log L)^{-1/8} }  \;=\;  (1.673 \pm 0.033) - 
   (1.056 \pm 0.098) {\log\log L \over \log L} 
 \label{eq5.7}
\ee 
with $\chi^2=2.76$ (3 DF, level = 43\%). However, the ratio 
$B/A \approx -0.63$ is quite different from the expected value 
$a'/(8 a^2) = -1/16 = -0.0625$.

Furthermore, if we try fitting
$\chi/ [L^{7/4} (\log L)^{-1/8}]$ to $A + B/\log L$,
we obtain an equally good (even slightly better) fit:
with the same $L_{min}=64$, we get
\be
{ \chi \over L^{7/4} (\log L)^{-1/8} }  \;=\; (1.454\pm0.013) - 
   { 0.600 \pm 0.055 \over \log L}    
 \label{eq5.8}
\ee
with $\chi^2=1.94$ (3 DF, level = 59\%).  
Clearly, at these modest values of $L$,
it is virtually impossible to disentangle numerically the
$\log\log L/\log L$ and $1/\log L$ contributions,
both of which are predicted theoretically to be present.
Indeed, the small value ($-1/16$) of the universal
$\log\log L/\log L$ coefficient makes it undetectable in the presence
of the much larger ($\sim -0.4$) nonuniversal $1/\log L$ coefficient.
To distinguish these two contributions, we would need to reach at least
$\log\log L \approx 5$, i.e.\ $L \approx 10^{64}$!
Note, finally, the large discrepancy between the estimates
\reff{eq5.7} and \reff{eq5.8} of the leading amplitude $A$:
it is virtually impossible to estimate the correct limiting value
in the presence of such strong corrections.

\subsubsection{Specific heat}

The FSS behavior of the specific heat at criticality is given by
\reff{scaling_cv_L};  it is of the same $L^p (\log L)^{-q}$ form
as the susceptibility, but with a much larger logarithmic exponent $q$.
This makes it extremely difficult to estimate accurately the leading exponent
$\alpha/\nu = 1$, as was found already in \cite{Salas_Sokal_AT}.

A naive power-law fit to $A L^{\alpha/\nu}$ gives a reasonably good 
result for $L_{min}=128$:
\be 
{\alpha \over \nu} = 0.770 \pm 0.008 
\ee 
with $\chi^2=1.15$ (2 DF, level = 57\%). This estimate is far below the 
exact value; this deviation is roughly consistent with the predicted  
multiplicative logarithmic correction 
in both sign and magnitude:
a behavior $(\log L)^{-3/2}$ can be well mimicked over the interval
$128 \le L \le 1024$ by a power $L^{-0.23}$.

We can try to estimate the power of the logarithmic term by 
fitting $C_H/L$ to $A \log^p L$. The result is good for $L_{min}=128$:
\be 
p = -1.258 \pm 0.044 
 \label{p_specheat}
\ee 
with $\chi^2=0.66$ (2 DF, level = 72\%). We obtain a much better result 
than for the susceptibility, but \reff{p_specheat} is still
six standard deviations away from the predicted value $-1.5$.
We can also try to compute $\alpha/\nu$ by fitting 
$C_H \log^{3/2} L$ to the power-law $A L^{\alpha/\nu}$. The result for 
$L_{min}=128$ is 
\be 
{\alpha \over \nu} = 1.044 \pm 0.008 
\ee
with $\chi^2=0.97$ (2 DF, level = 62\%). This is still five standard 
deviations away from the exact value $\alpha/\nu = 1$.

%
%

Finally, we have fitted the function $C_H/[L (\log L)^{-3/2}]$ to the Ansatz 
$A + B \log\log L/\log L$. The fit is good for $L_{min}=128$:
\be 
{C_H \over L (\log L)^{-3/2} } \;=\;  (1.607 \pm 0.113)  \,-\, 
      (1.926\pm 0.354) {\log\log L \over \log L}
\ee
with $\chi^2=0.72$ (2 DF, level = 70\%). Again the ratio $B/A \approx -1.2$ is
rather different from the expected universal value 
$3 a'/(2 a^2) = -3/4 = -0.75$ from \reff{scaling_cv_L}.

If, instead, we try to fit  $C_H/[L (\log L)^{-3/2}]$
to the Ansatz $A + B/\log L$, 
the fit is reasonable already for $L_{min}=64$:
\be
{C_H \over L (\log L)^{-3/2} } \;=\;  (1.291 \pm 0.022)  \,-\,
      {1.526 \pm 0.096  \over \log L}
\ee
with $\chi^2=1.22$ (3 DF, level = 75\%).
Again, we are unable to disentangle the two types of corrections.

\subsubsection{The combination $\chi/C_H^{1/12}$}

{}From \reff{scaling_cv_L}/\reff{scaling_chi_L} we see that the 
combination $\chi/C_H^{1/12}$ does not show any multiplicative logarithmic 
correction, nor any additive correction of order $\log\log L/\log L$.
Rather,
\be 
{\chi \over C_H^{1/12}}(0,0,\psi,L^{-1}) \sim  L^{5/3} \left[ 
   1 + O\!\left( {1 \over \log L} \right) \right] 
\label{scaling_ratio_L} 
\ee
It is therefore of some interest to see whether the fits for this
particular combination are at all ``cleaner'' than those for
$\chi$ and $C_H$ separately.
To compute the error bars on $\chi/C_H^{1/12}$, we took into account the 
cross-correlations between the specific heat and the susceptibility. 
We can thus trust $\chi^2$ value in these the fits.

To test the prediction \reff{scaling_ratio_L},
we first tried a power-law Ansatz $A L^p$.
The fit is already good for $L_{min}=16$:
\be 
p \;=\; 1.682 \pm 0.001 
\ee
with $\chi^2=3.60$ (5 DF, level = 61\%). This exponent is 15 standard deviations
away from the expected value $5/3 = 1.666667$. This discrepancy might be due to
a rather strong correction to scaling.

If we assume that the corrections to scaling are of order $1/\log L$, 
the fit is reasonable for $L_{min}=64$:
\be 
{\chi C_H^{-1/12} \over L^{5/3} } \;=\; (1.422\pm 0.013)  \,-\, 
                                   { 0.429 \pm 0.057 \over \log L} 
\ee
with $\chi^2=2.23$ (3 DF, level = 53\%).  
However, a correction of order $\log\log L/\log L$ gives 
an even better fit: already with $L_{min}=32$ we have
\be
{\chi C_H^{-1/12} \over L^{5/3} } \;=\; (1.588\pm 0.026)  \,-\,
                (0.785 \pm 0.078) {\log\log L \over \log L}
\ee
with $\chi^2=2.37$ (4 DF, level = 67\%). 
Again, we find it impossible to distinguish clearly between these two 
corrections to scaling, even when we know on theoretical grounds that only 
$1/\log L$ corrections are present.

\subsection{Finite-size scaling off criticality} \label{sec_res_off_crit}

In this section we are going to test the FSS predictions for the 4-state Potts
model off criticality. In particular, from \reff{scaling_xi_fundamental_three} 
we see that 
\be 
{\xi(\phi,0,\psi,L^{-1}) \over L} =  
\xi\!\left( \phi f_{{3\over2},{3\over4}}(\log L;\psi),0,0,1\right) + 
O\!\left( {1 \over \log L} \right) \;. 
\label{scaling_xi_off_criticality} 
\ee 
The first argument on the right-hand side is equal (for large $L$) to 
\be 
\phi f_{{3\over2},{3\over4}}(\log L;\psi) = 
{\phi L^{3/2} \over (-a\psi \log L)^{3/4} } 
\left[ 1 + {a'\over a^2} {\log\log L \over \log L} + 
       O\!\left( {1 \over \log L} \right) \right] 
 \;.
\ee 
This means that if we plot $\xi^{(2)}/L$ as a function of 
$\phi L^{3/2} \log^{-3/4} L$, the points will collapse onto a single curve. 
Of course, the expected deviations of order $\log\log L/\log L$
may make this collapse less than perfect.\footnote{ 
   One could also plot $\xi^{(2)}/L$ as a function of 
   $\phi L^{3/2} (\log L)^{-3/4} [1 + (a'/a^2) \log\log L/\log L ]$, and 
   get only $1/\log L$ corrections. However, we have seen in the preceding 
   subsection that it is very hard to disentangle these two effects at 
   criticality. Thus, we expect no gain in introducing explicitly the 
   $\log\log L/\log L$ correction.  
}

Analogously, from \reff{scaling_cv_general_L}--\reff{scaling_chi_general_L}
we obtain similar predictions for the specific heat and the susceptibility:
\begin{eqnarray} 
{C_H(\phi,0,\psi,L^{-1}) \over L (\log L)^{-3/2} } &=& 
  \widetilde{F}_{C_H}\!\left( {\phi L^{3/2} \over (\log L)^{3/4}},0,0,1\right) 
 + O\!\left( {\log\log L \over \log L} \right) 
 \label{scaling_cv_off_criticality} \\[1mm]
{\chi(\phi,0,\psi,L^{-1}) \over L^{7/4}(\log L)^{-1/8} } &=& 
\widetilde{F}_{\chi}\!\left( {\phi L^{3/2} \over (\log L)^{3/4}},0,0,1\right) +
 O\!\left( {\log\log L \over \log L} \right)
 \label{scaling_chi_off_criticality} 
\end{eqnarray} 
where $\widetilde{F}_{C_H}$ and $\widetilde{F}_{\chi}$ are certain 
derivatives of the free energy $f_{\rm sing}$. Again, plotting 
the l.h.s.\ of 
\reff{scaling_cv_off_criticality} or \reff{scaling_chi_off_criticality} versus
$\phi L^{3/2} (\log L)^{-3/4}$ will collapse the 
corresponding data onto a single curve, modulo corrections 
$O(\log\log L/\log L)$.

The effect of the marginal scaling field $\psi$ on the FSS equations 
is thus threefold:
1) The specific heat and the susceptibility
      have a multiplicative logarithmic correction.
2) The scaled-temperature variable appearing on the r.h.s.\ of the 
      FSS equations is not merely $L^{1/\nu}(J-J_c)$, but has a
      multiplicative logarithmic correction $(\log L)^{-3/4}$
      [the same for all observables].
3) The corrections to finite-size scaling are not $O(L^{-\omega})$,
      but rather $O(\log\log L/\log L)$, which makes 
      the numerical analysis much harder.

\begin{figure}
\epsfxsize=400pt\epsffile{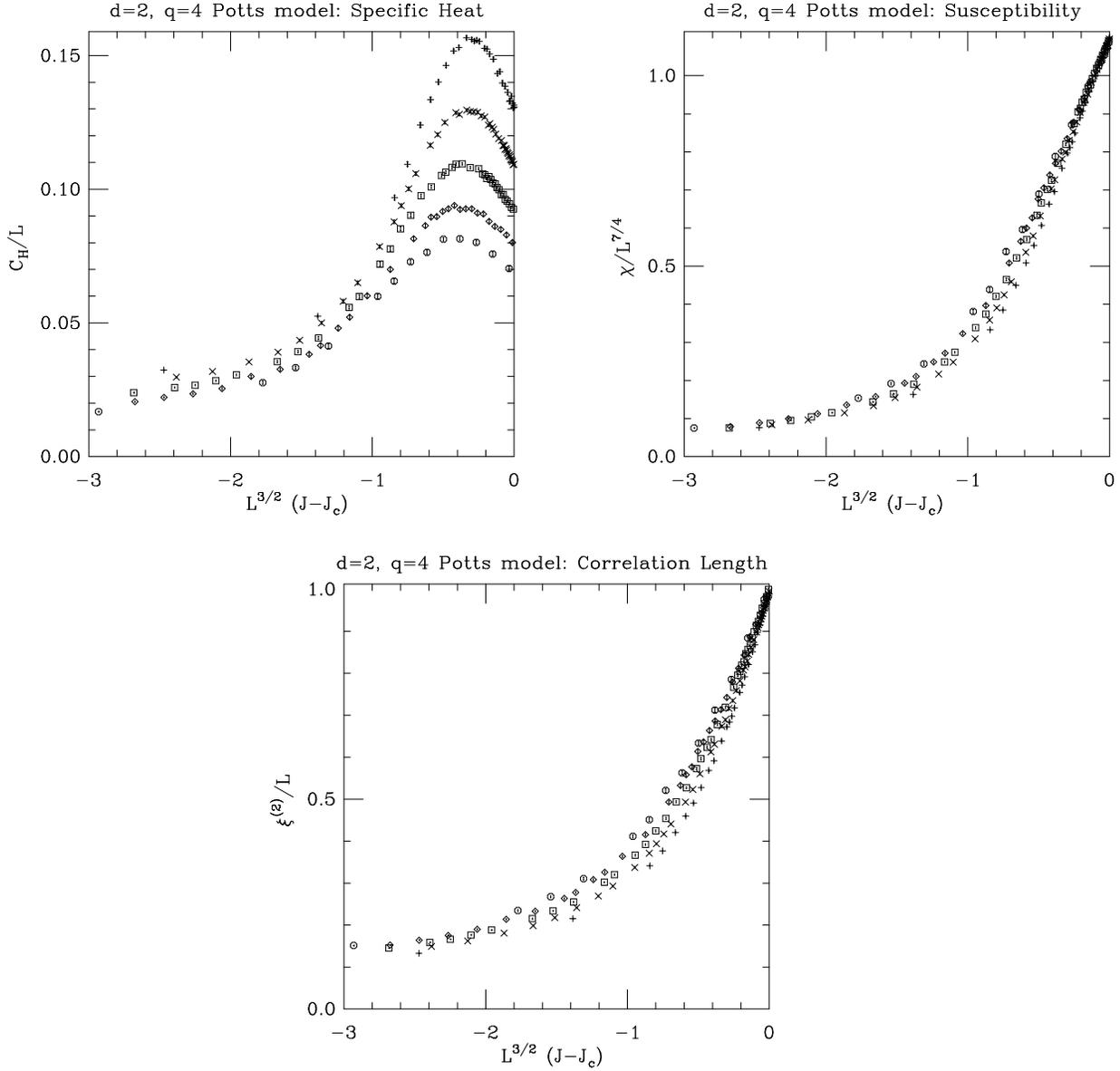}
  \caption{\protect\label{figure_fss_plot_vs_t}
           Naive finite-size scaling plots, neglecting all multiplicative 
           logarithms, for the specific heat $C_H$, the
           susceptibility $\chi$, and the second-moment correlation length 
           $\xi^{(2)}$. Note that he abscissa $L^{3/2}(J-J_c)$, where 
           $J_c = {1 \over 4}\log 3 = 0.274653\ldots\;$, also lacks the 
           predicted multiplicative logarithms. 
           Symbols denote the
           different lattice sizes: $L=32$ (+), $L=64$ ($\times$),
           $L=128$ ($\Box$), $L=256$ ($\Diamond$), and $L=512$ ($\circ$).
          }
\end{figure}

The naive approach to off-criticality FSS ---
neglecting all multiplicative logarithms ---
would be to plot $\xi^{(2)}/L$,
$C_H/L$, and $\chi/L^{7/4}$ versus $L^{3/2}(J-J_c)$. These plots are 
displayed in Figure~\ref{figure_fss_plot_vs_t}. The results are quite poor: 
there is no data-collapse for the specific heat; and for the susceptibility
and the correlation length the data collapse only close to $J_c$.   

\begin{figure}
\epsfxsize=400pt\epsffile{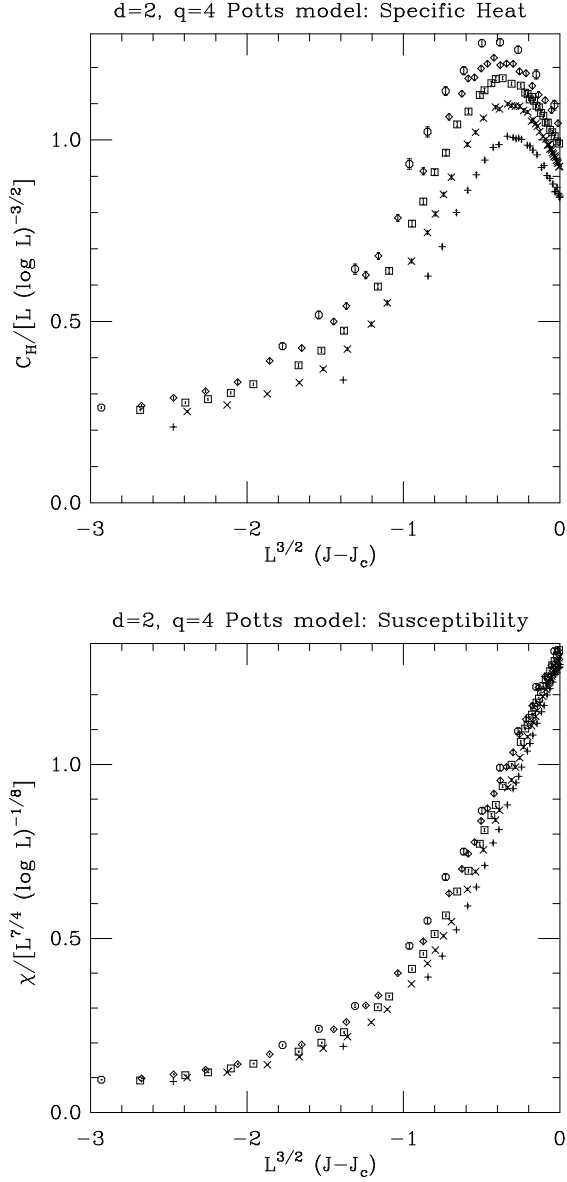}
  \caption{\protect\label{figure_fss_plot_vs_t2}
           Finite-size scaling plots that include the predicted multiplicative 
           logarithms $C_H \sim L (\log L)^{-3/2}$ and 
           $\chi \sim L^{7/4} (\log L)^{-1/8}$ but do {\em not}\/ include any 
           multiplicative logarithms in the abscissa $L^{3/2}(J-J_c)$. 
           Symbols denote the
           different lattice sizes: $L=32$ (+), $L=64$ ($\times$),
           $L=128$ ($\Box$), $L=256$ ($\Diamond$), and $L=512$ ($\circ$).
           Note that the fits are {\em worse}\/ than in 
           Figure~\protect\ref{figure_fss_plot_vs_t}.
          } 
\end{figure}

A slightly less naive approach is to incorporate the predicted
multiplicative logarithms for the specific heat and the susceptibility,
while still ignoring the multiplicative logarithmic
corrections to the abscissa $L^{3/2}(J-J_c)$.
But if we do this, we get even worse plots
(see Figure~\ref{figure_fss_plot_vs_t2}),
with the exception of the specific heat near $J_c$
(where the rescaling of the abscissa makes no difference anyway,
 and the multiplicative logarithm in $C_H$ helps a lot).
The slight deterioration in the susceptibility plot near $J_c$
is a reflection of our inability \reff{res_p_improved} 
to verify the correct power of $\log L$.

\begin{figure}
\epsfxsize=400pt\epsffile{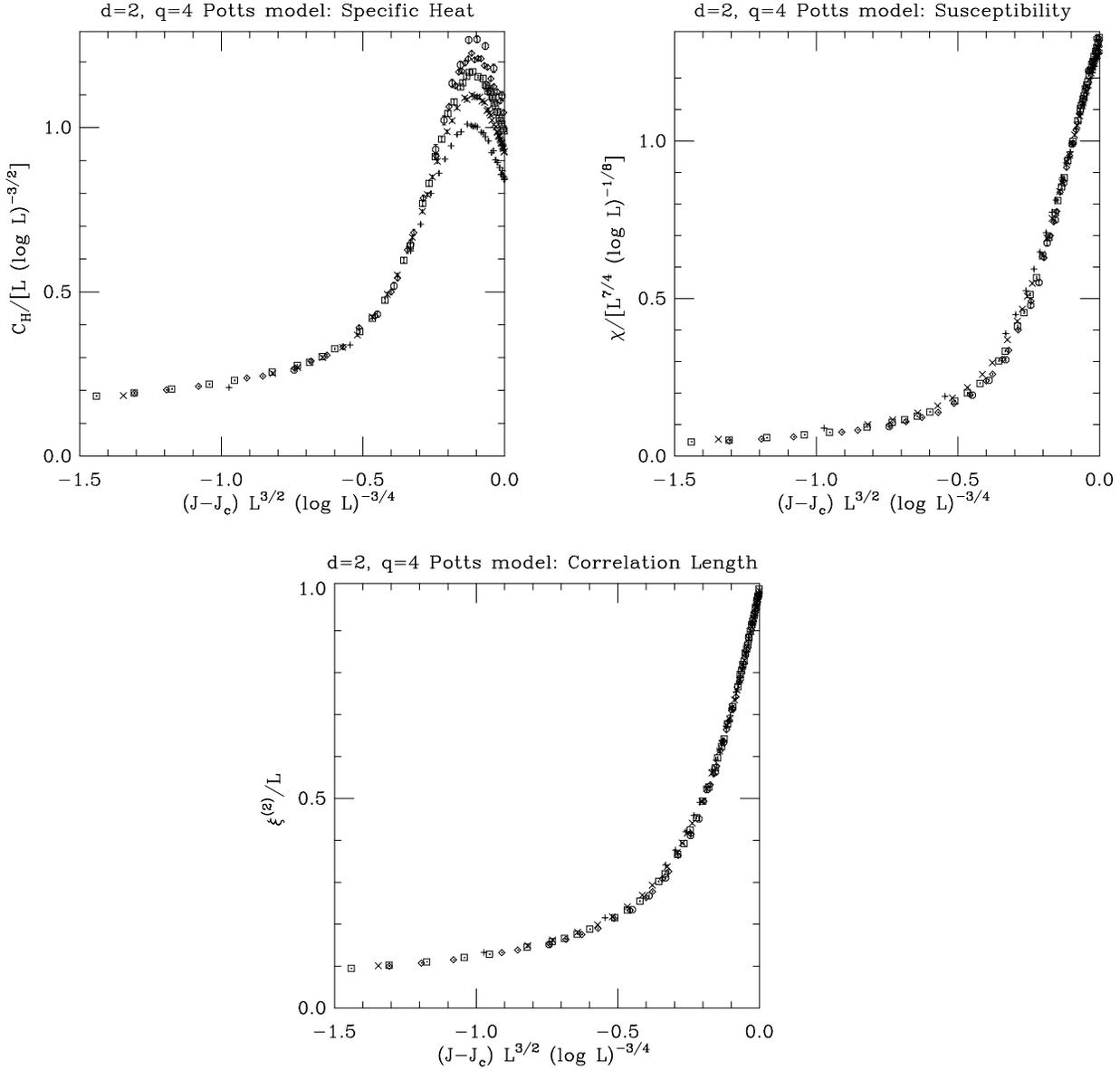}
  \caption{\protect\label{figure_fss_plot_vs_tG2}
           Finite-size scaling plots for the specific heat $C_H$, the
           susceptibility $\chi$, and the second-moment correlation length
           $\xi^{(2)}$ that take into account {\em all}\/ predicted 
           multiplicative logarithms. Note that the correct abscissa 
           variable is  
           $(J-J_c) L^{3/2}(\log L)^{-3/4}$. 
           Symbols denote the
           different lattice sizes: $L=32$ (+), $L=64$ ($\times$),
           $L=128$ ($\Box$), $L=256$ ($\Diamond$), and $L=512$ ($\circ$).
          }
\end{figure}

However, when we try to verify the correct FSS equations
\reff{scaling_xi_off_criticality}--\reff{scaling_chi_off_criticality}, the 
result is very different (see Figure~\ref{figure_fss_plot_vs_tG2}).
The data for the specific heat exhibit a good collapse away from $J_c$.
Close to the critical point we see large deviations,
presumably due to the $\log\log L/\log L$ and $1/\log L$ corrections.
For the susceptibility, we get a good (though not perfect)
data-collapse along the entire curve; the data-collapse is even better for the 
correlation length.
We conclude that the data are in reasonable agreement with the
predicted multiplicative logarithms, if we make allowance for
the additive logarithmic corrections to scaling.

\bigskip 

{\bf Remark.} In most situations, FSS plots of this type are difficult
to obtain because
of the uncertainties in the determination of $J_c$:
a small error on $J_c$ ruins the data-collapse.
Fortunately,
in our case we {\em do}\/ know the exact value of $J_c$. This fact allows 
us to test the FSS predictions very accurately.  

\bigskip

\begin{figure}
\epsfxsize=400pt\epsffile{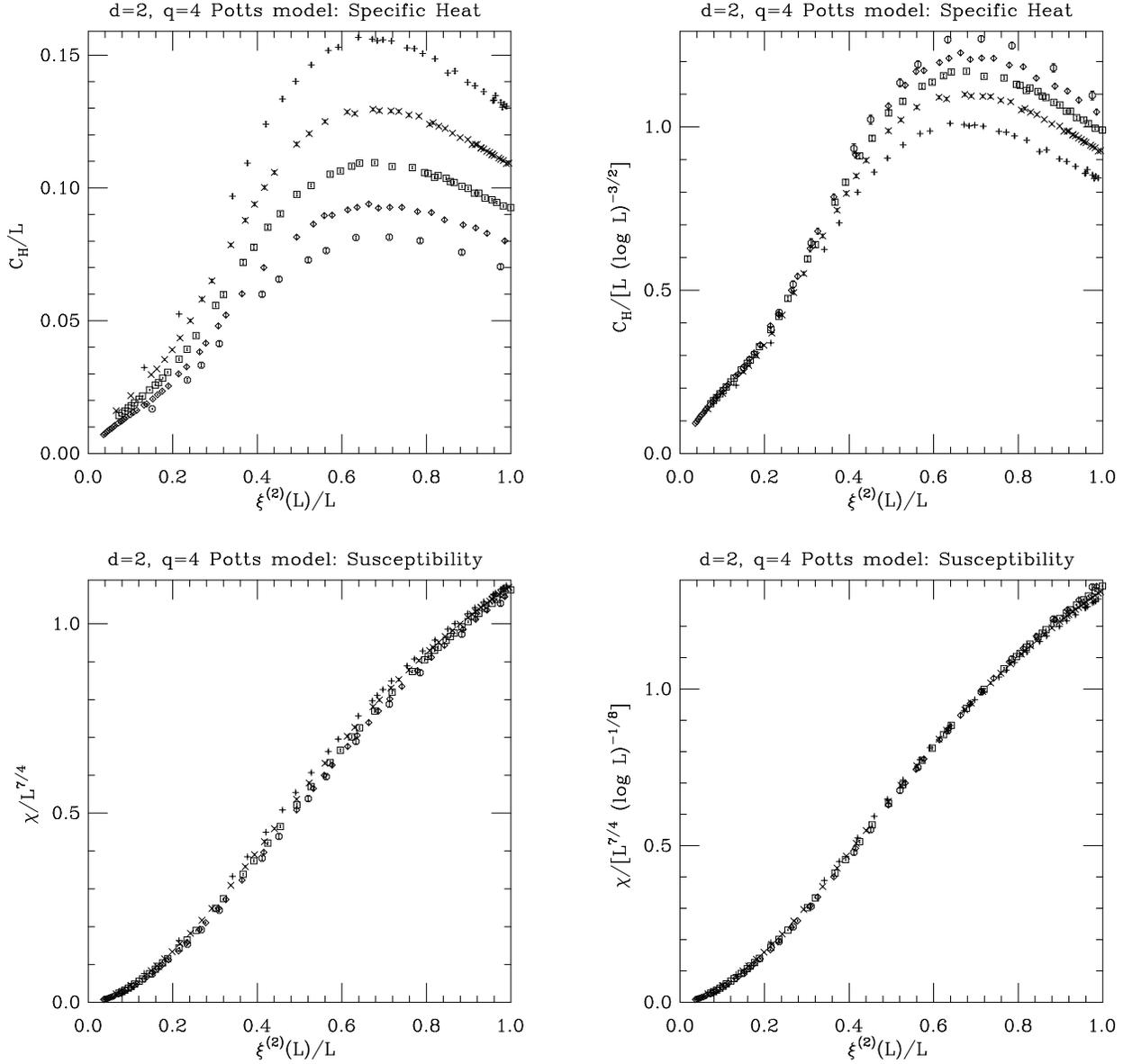}
  \caption{\protect\label{figure_fss_plot_vs_xi}
           Finite-size scaling plots for the specific heat $C_H$ and the
           susceptibility $\chi$ as functions of $\xi^{(2)}(L)/L$. 
           For each observable, the plot in the 
           left column includes only the leading power law, 
           while the plot in the right column takes into
           account the predicted multiplicative logarithmic correction.
           The improvement is clear.
           Symbols denote the
           different lattice sizes: $L=32$ (+), $L=64$ ($\times$),
           $L=128$ ($\Box$), $L=256$ ($\Diamond$), and $L=512$ ($\circ$).
          }
\end{figure}

Another way to present FSS data is to avoid using the
``bare'' variable $J-J_c$,
and to use instead the physical observable $\xi^{(2)}(L)/L$.
{}From \reff{scaling_cv_vs_xi}--\reff{scaling_chi_vs_xi}
we conclude that plotting 
$C_H/[ L (\log L)^{-3/2}]$ or $\chi/[L^{7/4} (\log L)^{-1/8}]$ versus 
$\xi^{(2)}/L$ will provide a single curve for each observable,
modulo corrections of order $\log\log L/\log L$. 
In Figure~\ref{figure_fss_plot_vs_xi} we display the FSS plots 
of these two observables, when we neglect the multiplicative logarithmic
corrections (left column) and with the full leading terms (right 
column).  
Including the multiplicative logarithmic corrections makes
a big improvement, especially for the specific heat.
Again we see large corrections to scaling near the 
critical point (namely, at $\xi^{(2)}/L \approx x^\star \approx 1.02$),
in agreement with the behavior found in Section~\ref{sec_res_crit}.

\begin{figure}
\epsfxsize=400pt\epsffile{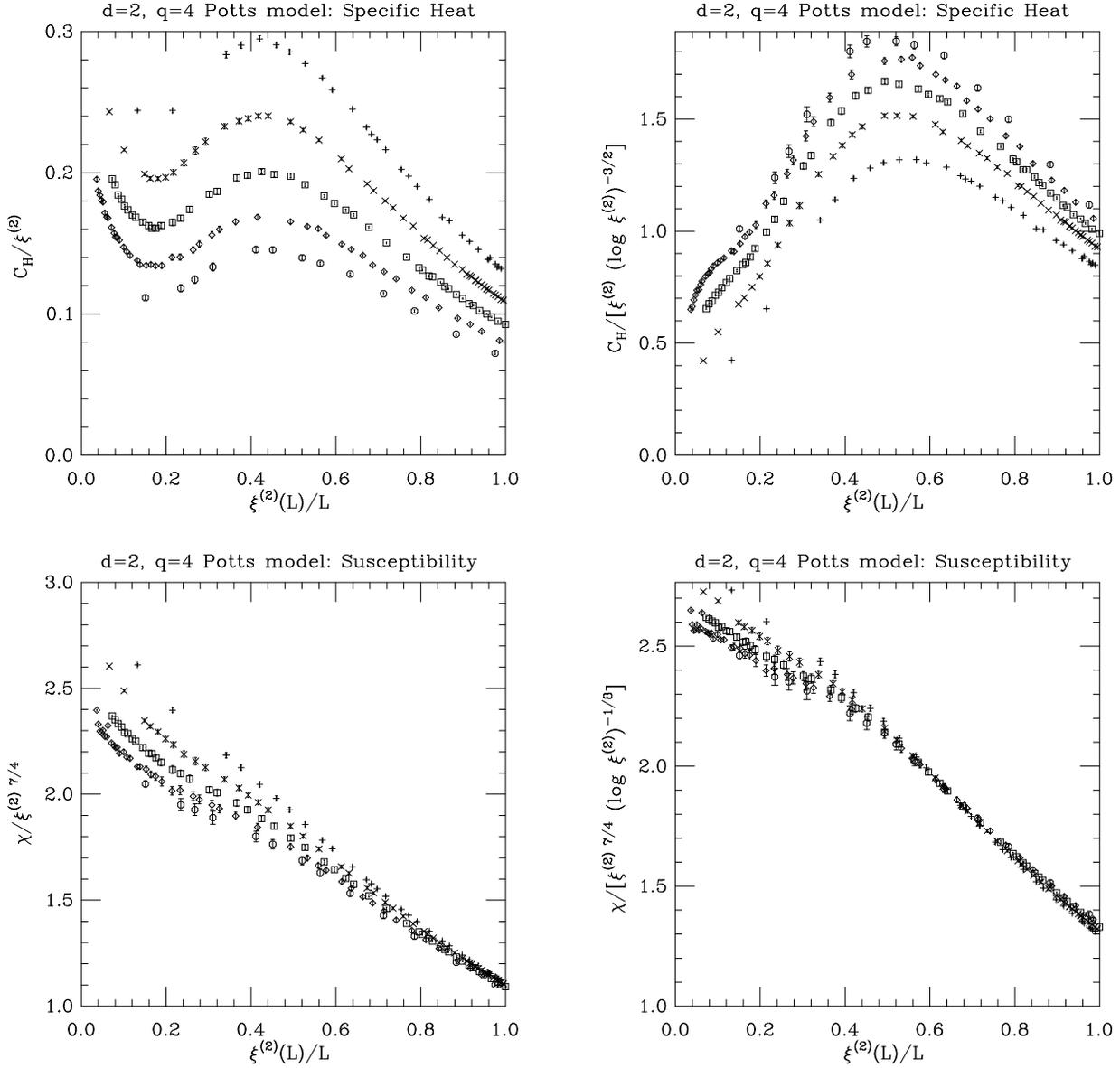}
  \caption{\protect\label{figure_fss_plot_vs_xi2}
           Finite-size scaling plots for the specific heat $C_H$ and the
           susceptibility $\chi$ as functions of $\xi^{(2)}(L)/L$.
           These are the same plots as in  
           Figure~\protect\ref{figure_fss_plot_vs_xi}, with the lattice size 
           $L$ replaced by the second-moment correlation length $\xi^{(2)}(L)$ 
           on the $y$-axis.  
           Symbols denote the
           different lattice sizes: $L=32$ (+), $L=64$ ($\times$),
           $L=128$ ($\Box$), $L=256$ ($\Diamond$), and $L=512$ ($\circ$).
          }
\end{figure}

The main drawback of this type of plot is that data with small 
$\xi^{(2)}/L$ are artificially compressed. One way to overcome this 
difficulty is to replace $L$ by $\xi^{(2)}$ in the $y$-axis. That is, we 
can plot $C_H/[ \xi^{(2)} (\log \xi^{(2)})^{-3/2}]$ and 
$\chi/[\xi^{(2) 7/4} (\log \xi^{(2)})^{-1/8}]$ versus 
$\xi^{(2)}/L$ (see Figure~\ref{figure_fss_plot_vs_xi2}). We emphasize that 
these plots are identical to those in Figure~\ref{figure_fss_plot_vs_xi},
except that the scale of the $y$-axis is changed in a way depending
on $x \equiv \xi^{(2)}/L$.
We see that the data-collapse at small $\xi^{(2)}/L$ is not in fact
as good as it had appeared in Figure~\ref{figure_fss_plot_vs_xi}:
in reality, it is mediocre for the susceptibility
and quite poor for the specific heat.
Nevertheless, there is a clear improvement in both cases
when we include the predicted multiplicative logarithmic correction.

\subsection{Extrapolation techniques} \label{sec_res_extra}

We have also considered how well the extrapolation scheme introduced in 
\cite{Sokal_extra} works in the presence of logarithmic corrections. 
This method makes use of the following FSS equation for an arbitrary 
long-distance observable ${\cal O}$:
\be 
{ {\cal O}(J,2L) \over {\cal O}(J,L)}  \;=\;   F_{\cal O}\!\left( 
                       {\xi(J,L) \over L} \right) + 
                       O(\xi^{-\omega},L^{-\omega}) \;,  
\label{scaling_extra_general} 
\ee
where $F_{\cal O}$ is an unknown scaling function and $\omega$ is a  
correction-to-scaling exponent.
(Here we have chosen a size-scaling factor $s=2$;
 a similar equation holds, of course, for any $s$.)

Let us first check the analogous equations for our case. To simplify the 
notation, we will write only two arguments for the observables: 
${\cal O}(J,L) \equiv {\cal O}(\phi,h=0,\psi,L^{-1})$.  
{}From \reff{scaling_xi_final} we conclude that 
\be 
{\xi(J,2L) \over \xi(J,L)}   \;=\;
   \widetilde{A}_\xi\!\left( {\xi_\infty(J) \over L} \right) + 
   \widetilde{B}_\xi\!\left( {\xi_\infty(J) \over L} \right) {1 \over \log L} 
  + \cdots   \;,
\ee
%
where $\widetilde{A}_\xi$ and $\widetilde{B}_\xi$ are universal 
(albeit unknown) scaling functions. 
Using the relation \reff{scaling_xi_infty_final} we arrive at 
\be
{\xi(J,2L) \over \xi(J,L)}   \;=\;
   A_\xi\!\left( {\xi(J,L) \over L} \right) +
   B_\xi\!\left( {\xi(J,L) \over L} \right) {1 \over \log L}
 + \cdots   \;,  
\label{scaling_extra_xi} 
\ee
where $A_\xi$ and $B_\xi$ are again universal.
%
%
Thus, we get the same equation as \reff{scaling_extra_general}, but the 
corrections to scaling are much larger ($\sim 1/\log L$).

The same type of equation can be derived
for the specific heat and the susceptibility,
using \reff{scaling_cv_vs_xi}--\reff{scaling_chi_vs_xi}:
although there are corrections of order $\log\log L/\log L$,
these cancel out in forming the ratio between
lattice sizes $L$ and $2L$. 
Notice that the nonuniversal $k_{\cal O}/\log L$ contribution also cancels out
in the ratio ${\cal O}(J,2L)/{\cal O}(J,L)$. This means that the leading
correction to scaling is universal. 
That is, we have
\be
{{\cal O}(J,2L) \over {\cal O}(J,L)}  =
   A_{\cal O}\!\left( {\xi(J,L) \over L} \right) + 
   B_{\cal O}\!\left( {\xi(J,L) \over L} \right) {1 \over \log L} + \cdots  
\label{scaling_extra_general_potts4}
\ee
%
%
where both scaling functions $A_{\cal O}$ and $B_{\cal O}$ are
{\em universal}\/.

%

\begin{figure}
\epsfxsize=400pt\epsffile{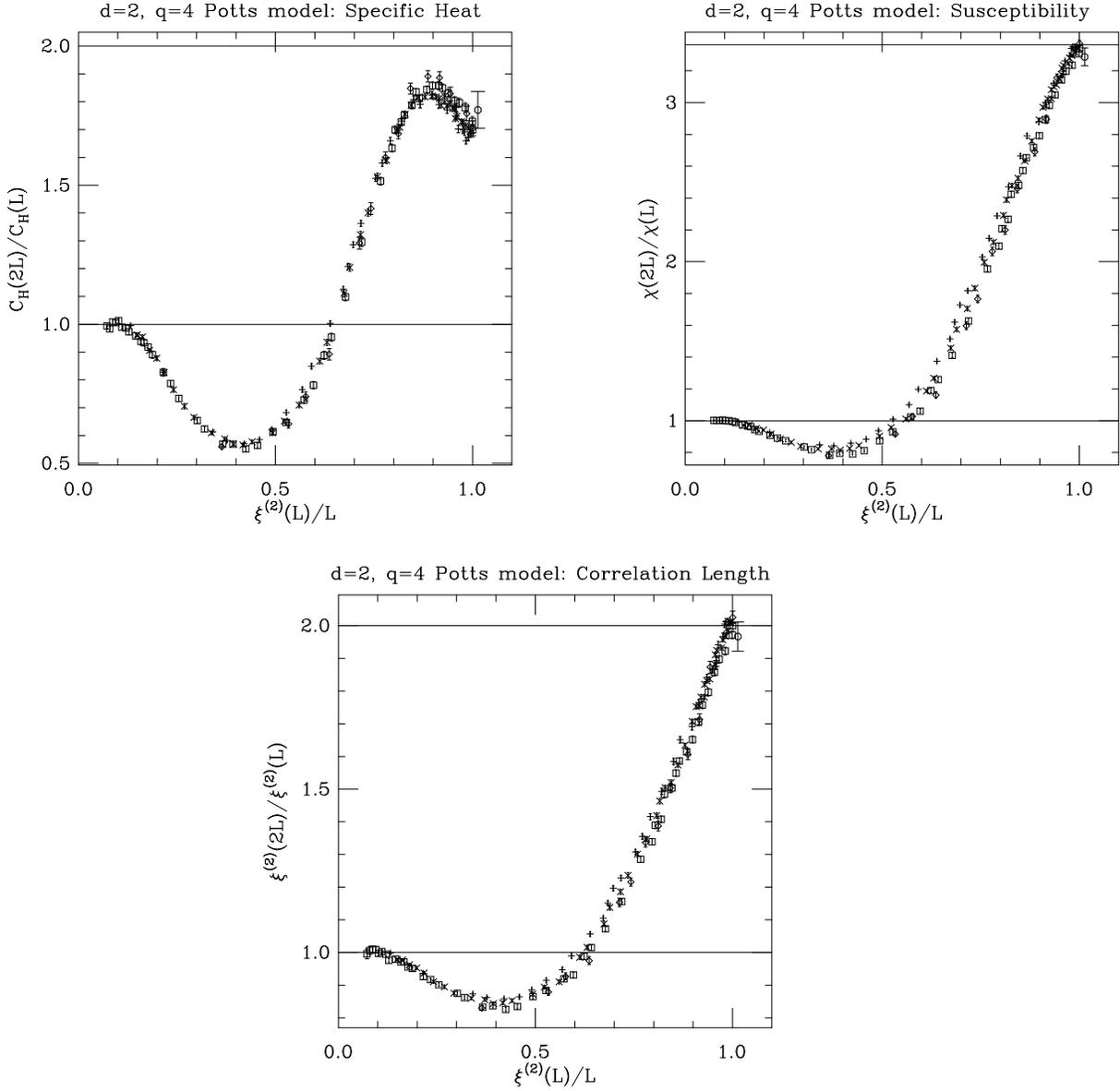}
  \caption{\protect\label{figure_plot_estra_std}
      ``$L/2L$'' finite-size scaling plots,
      of the kind employed in Ref.~\protect\cite{Sokal_extra},
      for the specific heat $C_H$, the  
           susceptibility $\chi$ and the second-moment correlation length 
           $\xi^{(2)}$.
           For each observable ${\cal O}$, 
           we plot the quantity ${\cal O}(2L)/{\cal O}(L)$ versus 
           $\xi^{(2)}/L$.  
           Symbols denote the
           different lattice sizes: $L=32$ (+), $L=64$ ($\times$),
           $L=128$ ($\Box$), $L=256$ ($\Diamond$), and $L=512$ ($\circ$).
          }
\end{figure}

The $1/\log L$ corrections in \reff{scaling_extra_general_potts4} 
make a numerical determination of the 
finite-size-scaling functions $A_{\cal O}$ very difficult.
Unfortunately, these functions are the basic objects
needed to extrapolate from 
finite $L$ to the infinite-volume limit \cite{Sokal_extra}. Thus, a naive 
application of the method of Ref.~\cite{Sokal_extra} is not expected to 
work well. Figure~\ref{figure_plot_estra_std} 
shows the plots of ${\cal O}(J,2L)/{\cal O}(J,L)$ versus $\xi^{(2)}/L$
for the specific heat, 
the susceptibility and the second-moment correlation length.
The data collapse is notably inferior to that observed in most other models
\cite{Sokal_extra}.
Note also that the specific-heat curve is {\em very far}\/ from its correct
value $2^{\alpha/\nu} = 2$ at the critical point
$\xi^{(2)}/L = x^\star \approx 1.02$;
rather, it takes the value
$2^{(\alpha/\nu)_{\rm eff}} = 2^{\approx 0.770} \approx 1.71$.
This is another indication of the very strong corrections to scaling.

\section{Dynamic critical behavior of the Swendsen-Wang-type algorithm}
\label{sec6}

In this section we shall analyze briefly the dynamic critical behavior
of our Swendsen-Wang-type algorithm
for the two-dimensional 4-state Potts model.
Recall that we have used the ``embedding'' variant of the SW-type
algorithm for the Ashkin--Teller model \cite{Salas_Sokal_LAT95,Salas_Sokal_AT}.
This algorithm does {\em not}\/ coincide with the standard
Swendsen-Wang \cite{Swendsen_87} algorithm for the 4-state Potts model,
but it is expected heuristically
(and confirmed numerically \cite{Salas_Sokal_AT} at criticality)
to lie in the same dynamic universality class.
We therefore expect the dynamic critical exponents to be identical
for the two algorithms;
and we also expect the dynamic finite-size-scaling functions to be identical
modulo a multiplicative factor.

\subsection{Dynamic finite-size scaling at criticality}

We first fit the integrated autocorrelation time for the energy,
$\tau_{{\rm int},{\cal E}}$, to a pure power law 
$\tau_{{\rm int},{\cal E}} = A L^{z_{{\rm int},{\cal E}}}$.
We obtain a good fit for $L_{min}=32$:
\be 
z_{{\rm int},{\cal E}} \;=\; 0.876 \pm 0.011
\ee
with $\chi^2=2.54$ (4 DF, level = 64\%). 
However, this {\em cannot}\/ be the true asymptotic
behavior, because the Li--Sokal bound \cite{Li_Sokal,Salas_Sokal_AT}
guarantees that $\tau_{{\rm int},{\cal E}} \ge {\rm const} \times C_H$,
and we know from \reff{scaling_cv_L} that the specific heat diverges
at criticality like $L (\log L)^{-3/2}$.\footnote{
   The Li--Sokal bound has been proven for the original
   Swendsen--Wang algorithm \cite{Li_Sokal} and more generally
   for the {\em direct}\/ form of the Ashkin--Teller SW algorithm
   \cite{Salas_Sokal_AT}.  But the embedding form of the AT algorithm
   is believed (and observed numerically \cite{Salas_Sokal_AT})
   to be in the same dynamic universality class as the direct algorithm.
   So the bound should apply for it as well.
}
Rather, this {\em apparent}\/ exponent $z_{{\rm int},{\cal E}} < 1$
is probably an effect of multiplicative logarithmic corrections
(with a negative exponent), just as it is for the specific heat.

In \cite{Salas_Sokal_AT,Salas_Sokal_Potts3} we 
showed that there are only two likely behaviors for the autocorrelation time
of the Swendsen--Wang algorithm for the 2D Potts models: 
\be
\tau_{{\rm int},{\cal E}}/C_H   \;=\;
 \cases{  A L^p  \quad\hbox{with $p$ small} \cr  
          A + B \log L  \cr
       }
\label{ansatz} 
\ee
With the available statistics it is very hard to distinguish between these 
two scenarios. If we fit the quantity 
$\tau_{{\rm int},{\cal E}}/[L (\log L)^{-3/2}]$ to a pure power law 
$A L^p$, we obtain a reasonable fit for $L_{min}=128$:
\be 
p \;=\; 0.153 \pm 0.028 
\ee
with $\chi^2=1.30$ (2 DF, level = 52\%). On the other hand, if we fit the 
same quantity to the logarithmic Ansatz $A + B \log L$, the fit is very 
good already for $L_{min}=16$:
\be
   {\tau_{{\rm int},{\cal E}}   \over   L (\log L)^{-3/2} }
   \;=\;
   (0.197 \pm 0.173)  \,+\,  (1.278 \pm 0.044) \log L
\ee
with $\chi^2=1.53$ (5 DF, level = 77\%).  
Thus, our MC data supports slightly better the logarithmic scenario compared to 
the power-law scenario. 
%
%

One can also study directly the ratio $\tau_{{\rm int},{\cal E}}/C_H$, and 
fit the results to the Ans\"atze \reff{ansatz}. Unfortunately, we do not know
the correct error bar on this ratio, as we do not know the 
covariance between the estimator of the specific heat and
the estimator of $\tau_{{\rm int},{\cal E}}$. Instead, 
we shall use the {\em upper bound}\/ on the error bar provided by the
triangle inequality.
Our error bars are thus overestimated --- by how much we do not know ---
and, consequently, the $\chi^2$ values of the fits are expected to be
artificially small.
We can, however, compare the {\em relative}\/ $\chi^2$ values for the two
Ans\"atze.
(See \cite{Salas_Sokal_AT} for a complete discussion concerning these points.)
If we try to fit $\tau_{{\rm int},{\cal E}}/C_H$
to a pure power law $A L^p$, we get a good result for $L_{min}=16$:
\be 
p = 0.119 \pm 0.011  
\ee 
with $\chi^2=1.06$ (5 DF, level = 96\%).
With the logarithmic Ansatz $A + B\log L$,
the fit is less good: for the same $L_{min}$ we 
obtain $\chi^2=1.99$ (5 DF, level = 85\%). Thus, for $L_{min}=16$
the power-law fit gives 
a $\chi^2$ a factor of two better than the logarithmic fit. However, for 
$L_{min}=32$, the two fits give more nearly equal values of $\chi^2$:
0.98 and 1.27, respectively.

\subsection{Dynamic finite-size scaling off criticality}

The FSS behavior of the dynamic observables
off criticality is expected to be similar to that discussed 
previously for the static observables. That is, we expect multiplicative 
logarithmic corrections, as well as additive logarithmic corrections to 
scaling. In this section we have considered the two scenarios discussed 
in the previous subsection, with the aim of distinguishing between them.

\begin{figure}
\epsfxsize=400pt\epsffile{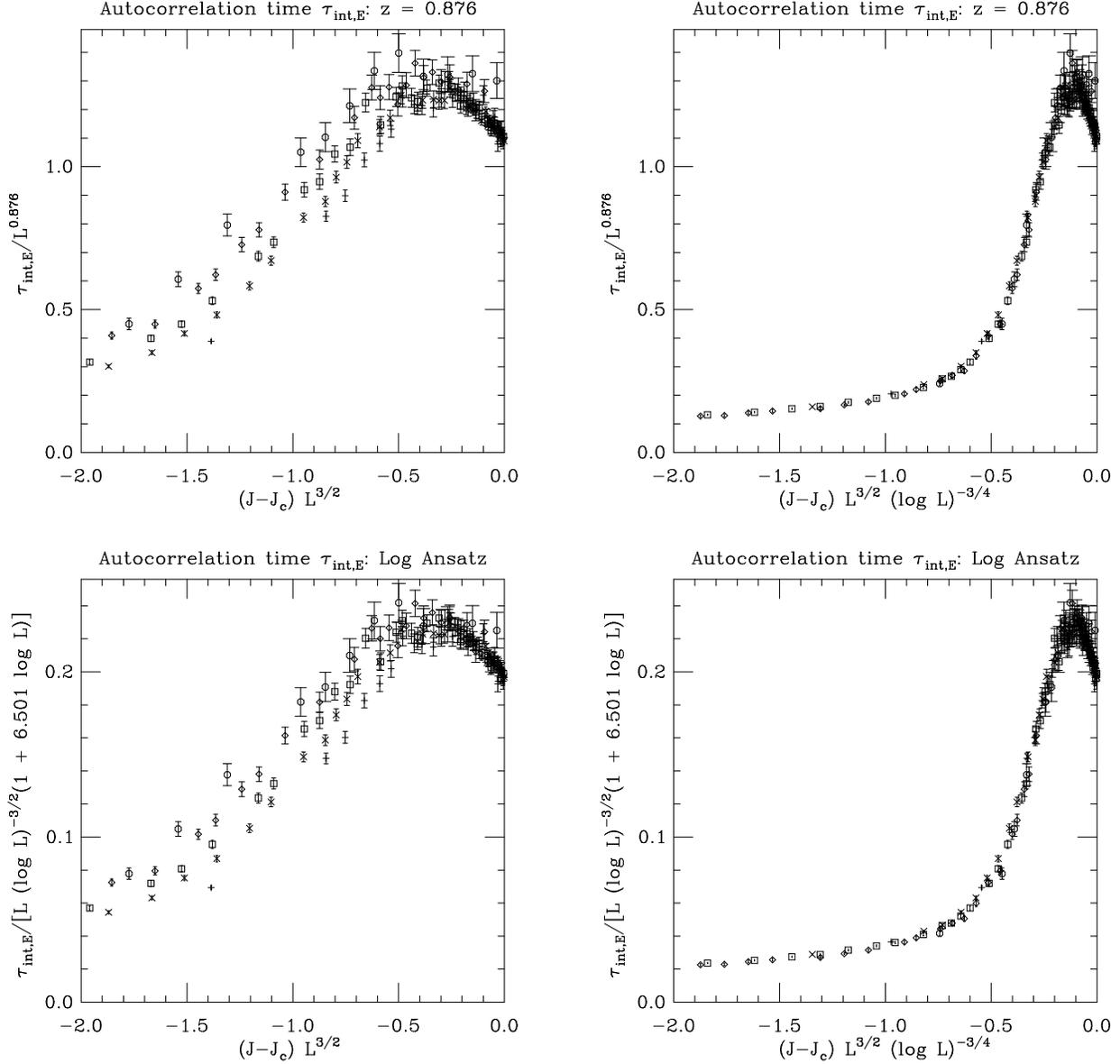}
  \caption{\protect\label{figure_fss_tauplot_vs_t}
           Finite-size scaling plots for the integrated autocorrelation time 
           $\tau_{{\rm int},{\cal E}}$ as a function of 
           $L^{3/2}(J-J_c)$ (left column) and $(J-J_c) L^{3/2}(\log L)^{-3/4}$ 
           (right column). We show the two most likely scenarios: 
           $\tau_{{\rm int},{\cal E}} \sim L^{z_{{\rm int},{\cal E}}}$
           with $z_{{\rm int},{\cal E}}=0.876\pm 0.011$ (upper row), and 
        $\tau_{{\rm int},{\cal E}} \sim L (\log L)^{-3/2}(1 + 6.501 \log L)$ 
           (lower row). 
           Symbols denote the
           different lattice sizes: $L=32$ (+), $L=64$ ($\times$),
           $L=128$ ($\Box$), $L=256$ ($\Diamond$), and $L=512$ ($\circ$).
          }
\end{figure}

First, we have made the FSS plots analogous to those of 
Figures~\ref{figure_fss_plot_vs_t} and \ref{figure_fss_plot_vs_t2}.
That is, we have plotted the leading behavior ---
$\tau_{{\rm int},{\cal E}}/L^{z_{{\rm int},{\cal E}}}$
or $\tau_{{\rm int},{\cal E}} / [L (\log L)^{-3/2}(1 + 6.501 \log L)]$ ---
versus $L^{3/2}(J-J_C)$ (left column in Figure~\ref{figure_fss_tauplot_vs_t}). 
In both cases, we obtain terrible fits. However, the situation improves 
dramatically when we consider the correct abscissa
$L^{3/2}(\log L)^{-3/4}(J-J_C)$ (right column in 
Figure~\ref{figure_fss_tauplot_vs_t}).  
Unfortunately, from these plots corresponding to the two most likely scenarios,
it is impossible to tell which one is more likely.

\begin{figure}
\epsfxsize=400pt\epsffile{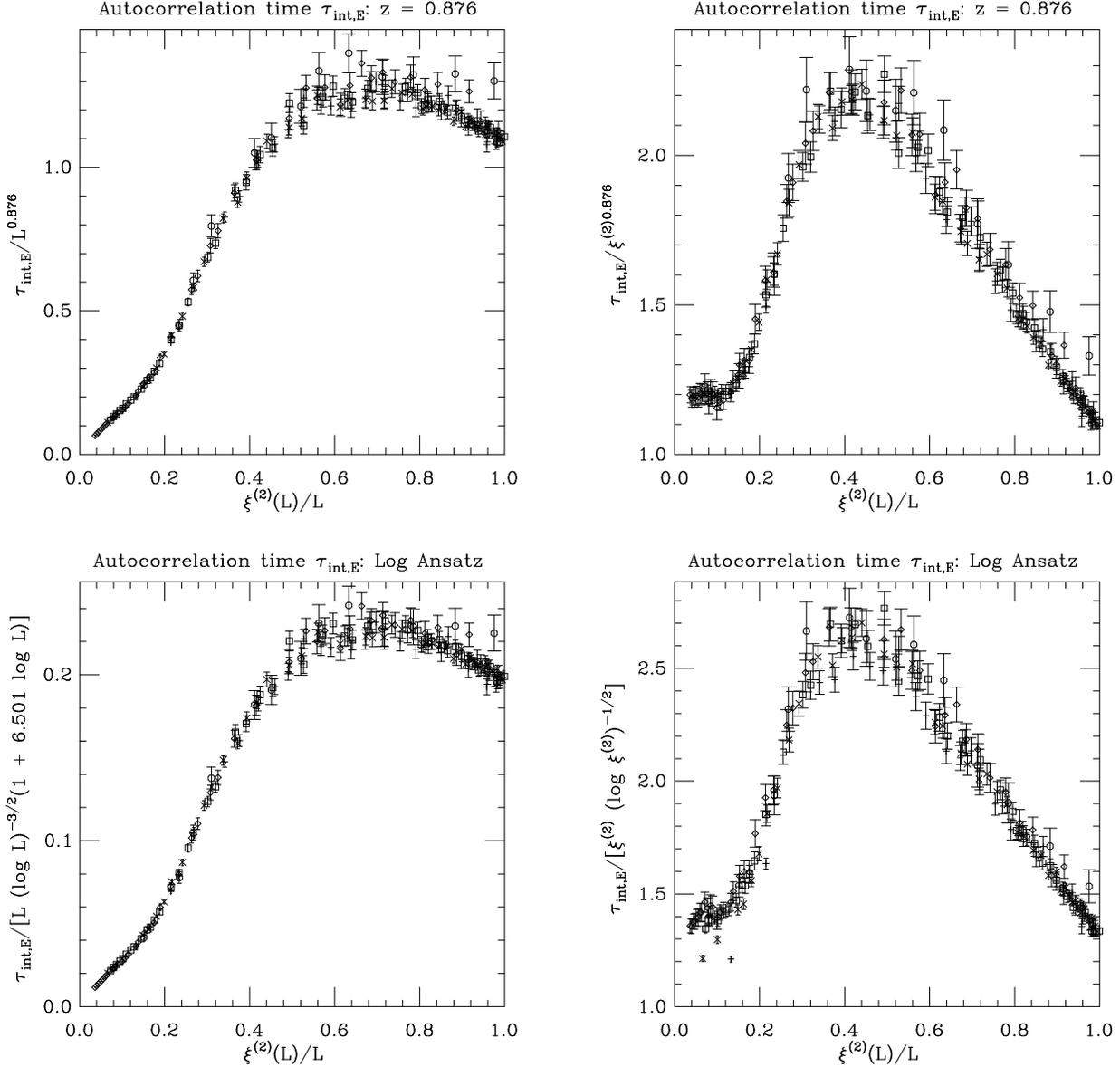}
  \caption{\protect\label{figure_fss_tauplot_vs_xi}
           Finite-size scaling plots for the integrated autocorrelation time
           $\tau_{{\rm int},{\cal E}}$ as a function of
           the physical observable
           $\xi^{(2)}(L)/L$. The upper row shows the Ansatz  
           $\tau_{{\rm int},{\cal E}} \sim L^{z_{{\rm int},{\cal E}}}$
           with $z_{{\rm int},{\cal E}}=0.876\pm 0.011$, while the 
           lower row shows the scenario  
        $\tau_{{\rm int},{\cal E}} \sim L (\log L)^{-3/2}(1 + 6.501 \log L)$. 
           In the left column we show the usual FSS plots; in the right 
           column we have substituted the variable $L$ in the $y$-axis
           denominator by $\xi^{(2)}$ in order to amplify the region with
           smaller $\xi^{(2)}/L$.  
           Symbols denote the
           different lattice sizes: $L=32$ (+), $L=64$ ($\times$),
           $L=128$ ($\Box$), $L=256$ ($\Diamond$), and $L=512$ ($\circ$).
          }
\end{figure}

Alternatively, we can plot the same quantities as functions of the 
physical observable $\xi^{(2)}(L)/L$ (left column in   
Figure~\ref{figure_fss_tauplot_vs_xi}). Again, the plots are very similar, 
and there is no objective reason to prefer one over the other. The 
corrections to scaling for small $\xi^{(2)}/L$ appear to be quite small;
they grow slightly in size as we approach larger values of $\xi^{(2)}/L$. 
To check that this behavior is not an artifact of the compression of the
vertical axis at small $\xi^{(2)}/L$,
we have tried replacing $L$ by $\xi^{(2)}$ in the $y$-axis denominators
(right column in Figure~\ref{figure_fss_tauplot_vs_xi}).
We emphasize that these plots are identical to those in the left column,
except that the scale of the $y$-axis is changed in a way depending
on $x \equiv \xi^{(2)}/L$.
Contrary to what we observed for 
the static observables,  the corrections to scaling for small 
$\xi^{(2)}/L$ are not so large compared to the error bars,
except for a few points at very small $\xi^{(2)}$
in the logarithmic Ansatz (where the neglect of the additive constant $A$ in
$A + B \log \xi^{(2)}$ obviously plays a role).
Note, finally, that the scaling function tends to a nonzero constant
when $\xi^{(2)}/L \to 0$, as expected on general theoretical grounds.

\appendix 

\section{Calculation of the coefficient $a'$}
\label{appendix_rg_flow} 

In this appendix we derive the value \reff{value_aprime}
of the cubic coefficient $a'$ in the RG flow for the 4-state Potts model.
The method is identical to that of Cardy {\em et al.}\/ \cite{Cardy_80},
carried to one higher order:
namely, we study the renormalization-group flow for the
$q$-state dilute Potts model in a neighborhood of the multicritical point
$q=4$, $\phi=h=\psi=0$;
we then match this flow with the exactly known \cite{Nienhuis_82}
values of the critical exponents as a function of $q$,
expanded now through {\em second}\/ order in $(4-q)^{1/2}$.

\subsection{Exactly known values of critical exponents}
\label{secA.1}

The leading thermal exponent $y_{T,1}$,
the next-to-leading thermal exponent $y_{T,2}$,
the leading magnetic exponent $y_{H,1}$
and the next-to-leading magnetic exponent $y_{H,2}$
of the $q$-state dilute Potts model are known exactly,
both for the ordinary critical point (arising in the pure Potts model)
and for the tricritical point (arising in the dilute model).
Their values are: 
\begin{eqnarray}
   y_{T,1}   & = &   {3+3x \over 2+x}             \label{value_yT1} \\[1mm]
   y_{T,2}   & = &   {4x   \over 2+x}             \label{value_yT2} \\[1mm]
   y_{H,1}     & = &   {15+8x+x^2   \over 4(2+x)}   \label{value_yH1} \\[1mm]
   y_{H,2}     & = &   {7 +8x+x^2   \over 4(2+x)}   \label{value_yH2}
\end{eqnarray}
where
\be
   x \;=\;  {2 \over \pi}  \arccos\!\left( { \sqrt{q} \over 2} \right)
   \;,
 \label{def_x}
\ee
with $-1 \le x \le 0$ corresponding to the ordinary critical points
and $0 \le x \le 1$ corresponding to the tricritical points.
The value of $y_{T,1}$ for the ordinary critical point was conjectured by
den Nijs \cite{denNijs_79}; this conjecture was extended to the 
tricritical branch of the dilute Potts model 
by Nienhuis {\em et al.}\/ \cite{Nienhuis_79}. 
It has now been derived by Coulomb-gas methods
\cite{Nienhuis_84,Nienhuis_DL11}
for both 
the critical \cite{Black_Emery} and tricritical \cite{Nienhuis_82} branches.  
The expression \reff{value_yT2} for the next-to-leading thermal exponent was  
first conjectured by Burkhardt \cite{Burkhardt_80}; it was derived using 
Coulomb-gas methods by Nienhuis \cite{Nienhuis_82}.   
Finally, the expressions \reff{value_yH1}/\reff{value_yH2}
for the leading and next-to-leading magnetic exponents
were first conjectured by Nienhuis {\em et al.}\/ \cite{Nienhuis_80} 
and Pearson \cite{Pearson_80};
their validity was established by 
den Nijs \cite{denNijs_83} using the Coulomb-gas approach. 
In the Coulomb-gas formulation, all these formulae are naturally parametrized
in terms of the variable $t = 1/(2+x)$,
which is proportional to the temperature of the associated Gaussian model;
the thermal (resp.\ magnetic) exponents are linear (resp.\ quadratic)
polynomials in $t$.

These exponents can also be understood in terms of conformal field theory
\cite{Belavin_84,Cardy_DL11,Ginsparg_89}.
Let us consider a conformal field theory (CFT) with central charge $c < 1$,
which we parametrize as
\be 
c \;=\; 1 \,-\, {6 \over m(m+1)}
\label{c_general}
\ee
with $0 < m < \infty$ (not necessarily integer).
Then the conformal weight of a primary field $\phi_{r,s}$
($r,s$ integer) is given by the Kac formula\footnote{
   We follow the notation of \cite{Cardy_DL11,Ginsparg_89}.
   In Ref.~\cite{Dotsenko_84a} the indices $r$ and $s$ are interchanged
   with respect to this notation.
}
\be
   \Delta_{r,s} \;=\; { [(m+1)r -m s]^2 - 1 \over 4 m (m+1) } \;.
\label{Kac_general}
\ee 
The corresponding critical exponent is $y_{r,s} = 2 - 2\Delta_{r,s}$.
If $m$ is a rational (resp.\ irrational) number, then the corresponding CFT
has a finite (resp.\ infinite) operator algebra.
If $m$ is an {\em integer}\/ $\ge 2$,
then the corresponding QFT is {\em unitary}\/,
at least as long as we restrict attention to the fields $\phi_{r,s}$
satisfying $1 \le s \le r \le m-1$.

The {\em critical}\/ $q$-state Potts model can be represented
\cite{Dotsenko_84a,Dotsenko_84b}
by a CFT with central charge 
\be 
   c  \;=\;  1  \,-\, {3x^2 \over 2 +x} \;,  
\label{c_critical} 
\ee 
where the parameter $x$
is given by \reff{def_x} with $-1 \leq x \leq 0$.\footnote{
   Our parameter $x$ is equal to minus the parameter $y$ appearing in 
   \cite{Dotsenko_84a,Cardy_DL11}. 
} 
We make the identification
\be 
x \;=\; -2/(m+1) \;,  
\ee
and the Kac formula \reff{Kac_general} reduces to 
\be 
\Delta_{r,s} \;=\; { [2(r-s) + sx]^2 - x^2 \over 8(2 +x) } \;.
\label{Kac_critical} 
\ee
The unitary theories $m=2,3,4,5,\ldots$ correspond to
$q=1,2,4\cos^2(\pi/5),3,\ldots\;$.

The energy operator $\varepsilon$ is identified with $\phi_{2,1}$. 
This means that $y_{T,1} = 2 - 2\Delta_{2,1}$, and we obtain the 
result \reff{value_yT1}. From the conformal algebra 
$\varepsilon\varepsilon \sim \phi_{2,1} \phi_{2,1} \sim I + \phi_{3,1}$,  
and we identify the operator $\phi_{3,1}$ as the one giving the next-to-leading 
thermal corrections. This implies that 
$y_{T,2} = 2 - 2\Delta_{3,1}$, and we get \reff{value_yT2}. 
Finally, the spin operator $s$ is identified with 
$\phi_{{m+1\over 2},{m+1\over 2}}$
(or alternatively $\phi_{{m+1\over 2}-1,{m+1\over 2}}$).
Therefore, 
$y_{H,1} = 2 - 2 \Delta_{{m+1\over 2},{m+1\over 2}} = 
2 - 2 \Delta_{{m+1\over 2},{m+1\over 2}-1}$,
and this yields \reff{value_yH1}.
The second magnetic exponent $y_{H,2}$ can be derived from the 
conformal weight of the operator $\phi_{{m+1\over 2}+1,{m+1\over 2}}$
(or alternatively $\phi_{{m+1\over 2}-2,{m+1\over 2}}$).
More generally, we can obtain the full spectrum of thermal exponents
\be
   y_{T,n}  \;=\;  2 - 2\Delta_{n+1,1}  \;=\;
                   {(4-n^2) + (n+2)x   \over  2+x}
\ee
and magnetic exponents
\be
   y_{H,n}  \;=\;  2 - 2\Delta_{{m+1\over 2}+n-1,{m+1\over 2}}  
            \;=\;  2 - 2\Delta_{{m+1\over 2}-n,{m+1\over 2}}
            \;=\;  {[16 - (2n-1)^2] +8x + x^2  \over  4(2+x)}
   \;.
\ee
However, this identification of the magnetic operators
makes sense only when $m$ is an odd integer $1,3,5,7,\ldots\,$,
corresponding to $q=0,2,3,2+\sqrt{2},\ldots\;$.

The {\em tricritical}\/ Potts model can be obtained by the
analytic continuation of the preceding formulae to $0 \leq x \leq 1$
\cite{Cardy_DL11}.
The conformal charge continues to be given by \reff{c_critical},
but we must now identify
\be 
x  \;=\; 2/m  \;.  
\ee
The Kac formula for the tricritical models is thus given by 
\be 
\Delta_{r,s} = { [2(r-s) + rx]^2 - x^2 \over 8(2 +x) } \;.
\label{Kac_tricritical}
\ee
The energy operator is now identified with the operator 
$\phi_{1,2}$, which reproduces \reff{value_yT1}.
More generally,
the thermal exponents are given by the operators $\phi_{1,1+n}$, while 
the magnetic exponents come from the operators 
$\phi_{{m\over2},{m\over2}+n}$ (or equivalently from 
$\phi_{{m\over2},{m\over2}+1-n}$).  Notice that the magnetic operators 
make sense only when $m$ is an {\em even}\/ integer $2,4,6,8,\ldots\,$, 
corresponding to the {\em same}\/ series $q=0,2,3,2+\sqrt{2},\ldots$
as in the critical case.

\subsection{Renormalization-group flow for the $q$-state dilute Potts model}

Following Cardy {\em et al.}\/ \cite[Section II]{Cardy_80},
we study the RG flow for the $q$-state dilute Potts model
in a neighborhood of the multicritical point $\epsilon \equiv q-4 =0$,
$\phi = h = \psi = 0$.
Cardy {\em et al.}\/ showed that the RG equations can be brought
by a series of smooth changes of variable into the form
\begin{subeqnarray}
   {d\psi \over dl} &=& a (\psi^2 + \epsilon)
           \,+\, O(f^3, \epsilon f, \epsilon^2)  \\
   {d\phi \over dl} &=& (y_T + b \psi) \phi
           \,+\, O(f^3, \epsilon f, \epsilon^2)  \\
   {d h   \over dl} &=& (y_H + c \psi) h
           \,+\, O(f^3, \epsilon f, \epsilon^2)
\label{rg_equations_general_q}
\end{subeqnarray}
where $a,b,c,y_T,y_H$ are universal parameters.
Here $f$ stands collectively for the fields $\psi$, $\phi$ and $h$;
and we consider $\epsilon$ to be of order $\psi^2$
because that is its magnitude at the fixed points
$\psi_\pm^2 = -\epsilon + O(\epsilon^2)$.

By a similar argument one can treat the terms of cubic order in the fields;
as before, the $\psi$ equation decouples from the others, and becomes
\be
   {d\psi \over dl}   \;=\;
   a (\psi^2 + \epsilon)  \,+\,  a' \psi^3  \,+\, a'' \epsilon \psi
       \,+\, O(\psi^4, \epsilon \psi^2, \epsilon^2)   \;.
\ee
For $\epsilon < 0$ there are fixed points at
\be
   \psi_\pm   \;=\;   \pm (-\epsilon)^{1/2}  \,+\,
                      {a'' - a'  \over 2a}  (-\epsilon)   \,+\,
                      O\bigl( (-\epsilon)^{3/2} \bigr)
\ee
with eigenvalues
\be
   y_\pm  \;=\;   \pm 2a (-\epsilon)^{1/2}  \,+\,
                      2a'  (-\epsilon)   \,+\,
                      O\bigl( (-\epsilon)^{3/2} \bigr)
   \;.
  \label{app_star2}
\ee
Here the $+$ sign corresponds to the tricritical point,
with a {\em relevant}\/ second thermal exponent $y_+ > 0$
controlling the tricritical-to-critical crossover;
and the $-$ sign corresponds to the ordinary critical point,
with an {\em irrelevant}\/ second thermal exponent $y_- < 0$
controlling the leading corrections to scaling.
Matching \reff{app_star2} to the known second thermal exponent
\reff{value_yT2}/\reff{def_x}, we find
\begin{eqnarray}
   a  & = &  {1 \over \pi}   \\[2mm]
   a' & = &  - {1 \over 2\pi^2} 
\end{eqnarray}
This reasoning gives an independent (and consistent!)\ derivation
of the quadratic coefficient $a$, which was derived previously by
Nauenberg and Scalapino \cite{Nauenberg_80,Cardy_80}
by matching (\ref{rg_equations_general_q}a,b)
to Baxter's \cite{Baxter_73} exact result for the latent heat
in the pure Potts model for $q>4$;
and it gives the promised derivation of $a'$.

{\bf Remark.}
This reasoning does not fix the value of $a''$.
Indeed, by a smooth change of variable
$\psi = \psi' + \alpha_2 \psi^{\prime 2}$,
the coefficient $a''$ can be set to {\em any}\/ desired value:
for example, $a''=0$ or $a''=a'$.  The latter choice has the property
of placing the fixed points at $\psi_\pm = \pm (-\epsilon)^{1/2}$
with {\em no}\/ correction $O(-\epsilon)$.
This approach can also be extended to higher order:
e.g.\ starting from the fourth-order equation
\begin{eqnarray}
   {d\psi \over dl}   & = &
   a (\psi^2 + \epsilon)  \,+\,  a' \psi^3  \,+\, a'' \epsilon \psi
                                                           \nonumber \\
       & & \qquad +\, a^{(3)} \psi^4  \,+\, a^{(4)} \epsilon \psi^2
                  \,+\, a^{(5)} \epsilon^2
       \,+\, O(\psi^5, \epsilon \psi^3, \epsilon^2 \psi, \epsilon^3)
\end{eqnarray}
we can make a smooth change of variable
$\psi = (1 + \alpha_{11} \epsilon) \psi' + \alpha_2 \psi^{\prime 2}
  + \alpha_3 \psi^{\prime 3}$
and fix {\em three}\/ of the four coefficients
$a'', a^{(3)}, a^{(4)}, a^{(5)}$ to any desired values.
For example, if we choose to fix $a''=a'$, $a^{(3)} = a^{(4)}$,
$a^{(5)} = 0$ we then have the normal form
\be
   {d\psi \over dl}   \;=\;  (\psi^2 + \epsilon) \widetilde{F}(\psi)
      \,+\, O(\psi^5, \epsilon \psi^3, \epsilon^2 \psi, \epsilon^3)
\ee
with $\widetilde{F}$ {\em independent}\/ of $\epsilon$.
Alternatively, we can choose $a''=a'$, $a^{(3)} = 0$, $a^{(4)} = a^{(5)}$
to yield the normal form
\be
   {d\psi \over dl}   \;=\;  (\psi^2 + \epsilon) 
       [A_0(\epsilon) + A_1(\epsilon) \psi]
      \,+\, O(\psi^5, \epsilon \psi^3, \epsilon^2 \psi, \epsilon^3)
\ee
with $A_0,A_1$ independent of $\psi$.
It would be interesting to know whether either (or both) of these normal forms
can be established to all orders in $\psi$ and $\epsilon$.
If so, the functions $\widetilde{F}(\psi)$, $A_0(\epsilon)$ and
$A_1(\epsilon)$ would then be {\em completely determined}\/
from the known exponent \reff{value_yT2}/\reff{def_x}.

%
%
\section*{Acknowledgments}

We wish to thank John Cardy for correspondence. 
The authors' research was supported in part 
by U.S.\ National Science Foundation grants DMS-9200719
and PHY-9520978, and by NSF Metacenter grant MCA94P032P
providing computing resources at the Pittsburgh Supercomputing Center and  
the Cornell Theory Center.

\newpage
\renewcommand{\baselinestretch}{1}
\large\normalsize
%
%
%
%

\clearpage

\begin{table} 
\centering 
\vspace*{-0.5in} 
\addtolength{\tabcolsep}{-1.0mm}
\begin{tabular}{|r|r|r@{$\,\pm\,$}r|r@{$\,\pm\,$}r|%
                     r@{$\,\pm\,$}r|r@{$\,\pm\,$}r|}%
\hline\hline  \\[-0.5cm] 
\multicolumn{1}{|c|}{$L$}    & 
\multicolumn{1}{c|}{$MCS$}   & 
\multicolumn{2}{c|}{$\chi$}  & 
\multicolumn{2}{c|}{$C_H$}   & 
\multicolumn{2}{c|}{$\xi^{(2)}$}   & 
\multicolumn{2}{c|}{$\tau_{{\rm int},{\cal E}}$} \\[0.1cm] 
\hline\hline 
   16 & 0.9 &   141.41 &   0.29 &     2.513 &   0.013 &    15.758 &   0.056 &    12.86 &   0.24 \\ 
   32 & 1.9 &   474.23 &   0.94 &     4.170 &   0.020 &    31.681 &   0.100 &    23.13 &   0.40 \\ 
   64 & 9.9 &  1587.70 &   1.90 &     6.971 &   0.020 &    63.704 &   0.115 &    41.58 &   0.42 \\ 
  128 & 6.9 &  5321.74 &  10.55 &    11.730 &   0.055 &   128.153 &   0.368 &    77.13 &   1.26 \\ 
  256 & 4.4 & 17764.70 &  60.96 &    19.936 &   0.158 &   256.291 &   1.243 &   143.00 &   3.99 \\ 
  512 & 2.9 & 59876.54 & 334.67 &    33.969 &   0.467 &   519.102 &   4.078 &   252.83 &  11.57 \\ 
 1024 & 0.8 & 196872.43 & 3137.60 &    60.135 &   2.091 &  1020.767 &  21.680 &   534.44 &  67.68 \\ 
[0.1cm] 
\hline\hline
\end{tabular}
\caption{Monte Carlo data for the 4-state Potts model at criticality. For each lattice size $L$ 
we show the number of measurements ($MCS$) in units of $10^5$, the susceptibility $\chi$, 
the specific heat $C_H$, the second-moment correlation length $\xi^{(2)}$, and the integrated 
autocorrelation time for the energy $\tau_{{\rm int},{\cal E}}$. The quoted error bars 
correspond to one standard deviation (i.e., confidence level $\sim 68\%$).}
\label{table_mc_data_tc}
\end{table}

\clearpage 

%
%
%
%
\newpage

\clearpage 

%
%
%
%
\newpage

\clearpage 

%
%
%
%
\newpage

\clearpage 

%
%
%
%
\newpage

\clearpage 

%
%
%
%
\newpage

%
%
%
%
\newpage

%
%
\newpage

%
%
\newpage
%
%
\end{document}